\documentclass[pra,aps,superscriptaddress,amssymb,amsmath,twocolumn,bibliography]{revtex4-2}
\usepackage{graphicx}
\usepackage{placeins}
\usepackage{hyperref}
\usepackage{xcolor}
\usepackage{listings}
\usepackage[normalem]{ulem}
\usepackage[utf8]{inputenc}
\usepackage{color}
\usepackage{float}
\usepackage{simplewick}
\usepackage{algorithm}
\usepackage{algorithmic}

\usepackage{tabu}
\usepackage{epstopdf}
\usepackage{bm}
\usepackage{xspace}

\begin{document}
	
\title{Improving the convergence of an iterative algorithm for solving arbitrary linear equation systems using classical or quantum binary optimization.}

\author{Erick R. Castro}
\email[]{erickc@cbpf.br}
\affiliation{Centro Brasileiro de Pesquisas F\'{\i}sicas,
	22290-180, Rio de Janeiro, RJ, Brazil} 

\author{Eldues O. Martins}
\email[]{eldues@petrobras.com.br}
\affiliation{Petróleo Brasileiro S.A., Centro de Pesquisas Leopoldo Miguez de Mello, Rio de Janeiro, Brazil} 
\affiliation{Centro Brasileiro de Pesquisas F\'{\i}sicas,
	22290-180, Rio de Janeiro, RJ, Brazil} 

\author{Roberto S. Sarthour}
\email[]{sarthour@cbpf.br}
\affiliation{Centro Brasileiro de Pesquisas F\'{\i}sicas,
	22290-180, Rio de Janeiro, RJ, Brazil} 

\author{Alexandre M. Souza}
\email[]{amsouza@cbpf.br}
\affiliation{Centro Brasileiro de Pesquisas F\'{\i}sicas,
	22290-180, Rio de Janeiro, RJ, Brazil}

\author{Ivan S. Oliveira}
\email[]{ivan@cbpf.br}
\affiliation{Centro Brasileiro de Pesquisas F\'{\i}sicas,
	22290-180, Rio de Janeiro, RJ, Brazil}

\begin{abstract}
Recent advancements in quantum computing and quantum-inspired algorithms have sparked renewed interest in binary optimization. These hardware and software innovations promise to revolutionize solution times for complex problems. In this work, we propose a novel method for solving linear systems. Our approach leverages binary optimization, making it particularly well-suited for problems with large condition numbers. We transform the linear system into a binary optimization problem, drawing inspiration from the geometry of the original problem and resembling the conjugate gradient method. This approach employs conjugate directions that significantly accelerate the algorithm's convergence rate. Furthermore, we demonstrate that by leveraging partial knowledge of the problem’s intrinsic geometry, we can decompose the original problem into smaller, independent sub-problems. These sub-problems can be efficiently tackled using either quantum or classical solvers. While determining the problem’s geometry introduces some additional computational cost, this investment is outweighed by the substantial performance gains compared to existing methods.
\end{abstract}	

\maketitle

\section{Introduction}

Quadratic unconstrained binary optimization problems (QUBO) \cite{Kochenberger2014} are equivalent formulations of some specific type of combinatorial optimization problems, where one (or a few) particular configuration is sought among a finite huge space of possible configurations. This configuration maximizes the gain (or minimizes the cost) of a real function $f$ defined in the total space of possible configurations. In QUBO problems, each configuration is represented by a binary $N$-dimensional vector $\mathbf{q}$ and the function $f$ to be optimized is constructed using a $N\times N$ symmetric matrix $\mathbf{Q}$. For each possible configuration, we have:  

\begin{equation}\label{QUBOFuncFinal}
f(\mathbf{q})=\mathbf{q}^{T}.\mathbf{Q}.\mathbf{q}.
\end{equation}

The sought optimal solution $\mathbf{q}^{*}$ satisfies $f(\mathbf{q}^{*})<\epsilon$, with $\epsilon$ a sufficiently small positive number. It is often easier to build a system configured near the optimal solution than to build a system configured at the optimal solution. 

QUBO problem is NP-Hard and is equivalent to finding the ground state of a general Ising model with an arbitrary value and numbers of interactions, commonly used in condensed matter physics \cite{Lucas2014,Barahona1982}. The ground state of the related quantum Hamiltonian encodes the optimal configuration and can be obtained from a general initial Hamiltonian using a quantum evolution protocol. This is the essence of quantum computation by quantum annealing \cite{Kadowaki1998}, where the optimal solution is encoded in an physical Ising quantum ground state. Hybrid quantum–classical methods, digital analog algorithms, and classical computing inspired by quantum computation are promising Ising solvers, see \cite{Mohseni2022}.

Essential classes of problems, not necessarily combinatorial, can be handled using QUBO-solvers. As an example, the problem of solving systems of linear equations was previously studied in references \cite{OMalley2016, Rogers2020, Pollachini2021,Souza2021}, in the context of quantum annealing. The complexity and usefulness of the approach were discussed in references \cite{Borle2019, Borle2022}. From those, we can say that quantum annealing is promising for solving linear equations even for ill-conditioned systems and when the number of rows far exceeds the number of columns.

 In another context, QUBO formulation protocols to train machine learning models were recently developed with the promising expectation that quantum annealing could solve this type of hard problem more efficiently \cite{Date2021}. Machine learning algorithms and specific quantum-inspired formulations of these strategies in the quantum circuit approach have grown substantially in recent years, see for example \cite{Hua2024,Gong2024,ChenGong2024,Zhou2023,Huang2023,Wu2022} and references therein. At the core of the machine learning approach, linear algebra is a fundamental tool used in these formulations. Therefore, the study of QUBO formulations of linear problems and their enhancement can be of interest in the use of the quantum annealing process in machine learning approaches. Another recent example is the study of simplified binary models of inverse problems where the QUBO matrix represents a quadratic approximation of the forward non-linear problem, see \cite{Greer2023}. It is interesting to note that in classical inverse problems, the necessity of solving linear system equations is an essential step in the whole process.

In this work we propose a new method to enhance the convergence rate of an iterative algorithm used to solve a system of equations with an arbitrary condition number. At each stage, the algorithm maps the linear problem to a QUBO problem and finds appropriate configurations using a QUBO solver, either classical or quantum. In previous implementations, the feasibility of the method is linked to the specific binary approximation used. Generally, as the condition number increases, more bits are required, which increases the dimension of the QUBO problem. Our contribution shows that a total or partial knowledge of the intrinsic geometry of the problem helps to reformulate the QUBO problem stabilizing the convergence to the solution and therefore improving the performance of the algorithm. In the case of full knowledge of the geometry, we show that the associated QUBO problem is trivial. If the geometry is only partially solved, we show that the QUBO problems to solve are small and, in principle, attainable with low binary approximation.

The paper is organized as follows: section \ref{Sec1} briefly describes how to convert the problem of solving a system of linear equations in a QUBO problem. The conventional algorithm for this problem is presented and illustrated with examples. Subsequently, we analyze the geometrical structure of the linear problem $\mathbf{A}\cdot\mathbf{x} = \mathbf{b}$ and their relation with the function (\ref{QUBOFuncFinal}); from them, a new set of QUBO configurations is proposed, attending the intrinsic geometry in a new lattice configuration. In section \ref{Sec2}, we implement these ideas in a new algorithm using a different orthogonality notion (that we call $\mathbf{H}$-orthogonality) related to the well-known gradient descent method. Using the $N\times N$ matrix $\mathbf{A}$, we find a new set of $N$ vectors that characterize the geometry of the problem. We compare the new algorithm with the previous version revised in section \ref{Sec1}. Section \ref{Sec3} uses the tools of the previous section to construct a different set of vectors grouped in many subsets mutually $\mathbf{H}$-orthogonal. This construction allows the decomposing of the original QUBO problem into independent QUBO sub-problems of smaller dimensions. Each sub-problem can be addressed using quantum or classical QUBO-solvers, allowing arbitrary linear equation systems to be resolved. In section \ref{Sec4}, we present the final considerations. 

\section{System of linear equations}\label{Sec1}

\subsection{Writing a system of equations as a QUBO problem}\label{Sec1A}

Solving a system of $N$ linear equations of $N$ variables is identical to finding a $N$-dimensional vector $\mathbf{x}$ $\in$ $\mathbb{R}^N$ that satisfies

\begin{equation}
	\mathbf{A}\cdot \mathbf{x} = \mathbf{b},
\end{equation}
where $\mathbf{A}$ is the matrix constructed with the coefficients of the $N$ linear equations and $\mathbf{b}$ is the vector formed with the inhomogeneous coefficients. If the determinant $\mathrm{Det}(A) \neq 0$,  then there exists one unique vector $\mathbf{x}^*$ that solves the linear system. We can transform the linear problem of real variables into  a binary optimization problem using a binary $R$-approximation of the components of one vector $\hat{\mathbf{x}}$:

\begin{equation}\label{NormVectors}
	\hat{x}_i=\sum_{r=0}^{R-1}q_i^{(r)}2^{-r}.
\end{equation}

Defining the vector $\mathbf{q}^{(r)}=(q_1^{(r)},\cdots,q_N^{(r)})$. The relation between $\mathbf{x}$ and the binary numbers $q_i^{(r)}$ is

\begin{equation}
	\mathbf{x} = \mathbf{x}_0 + L\sum_{r=0}^{R-1}2^{-r}\left(\mathbf{q}^{(r)}-\frac{\mathbf{I}}{2}\right),
\end{equation}
where $L$ is the length of the edge of the $N$-cube and $\mathbf{I}$ is the $N$-vector $(1,1,\cdots,1)$. Utilizing equation (\ref{NormVectors}) and recognizing the summation involving the $\mathbf{I}$ term, we can express

\begin{equation}\label{Vectors}
	\mathbf{x} = \mathbf{x}_0 + L\hat{\mathbf{x}} - \frac{2^R-1}{2^R}L\mathbf{I},
\end{equation}
where $\hat{\mathbf{x}}=(\hat{x}_1,\cdots,\hat{x}_N)$. With this notation, each binary vector $$\mathbf{q}=(q_1^{(0)}\cdots,q_1^{(R-1)},q_2^{(0)},\cdots,q_2^{(R-1)},\cdots,q_N^{(R-1)})$$ of length $RN$ defines an unique vector $\mathbf{x}$. These choices ensure that the initial guess $\mathbf{x}_0$ remains at the center of the $N$-cube.

To construct the QUBO problem associated with solving the linear system, we provide a concrete example with $N=2$; the generalization to arbitrary $N$ is straightforward. Let $\mathbf{A}$ be the matrix and $\mathbf{b}$ be the vector.

\begin{equation}
\mathbf{A}=\begin{pmatrix}
1 & 2 \\ 3 & 4
\end{pmatrix}, \,\,\,\,\,\,\,\, \mathbf{b}=\begin{pmatrix}
5 \\ 6
\end{pmatrix}.\label{LS1}
\end{equation}

\begin{figure*}
 	\centering
 	\includegraphics[width=2\columnwidth]{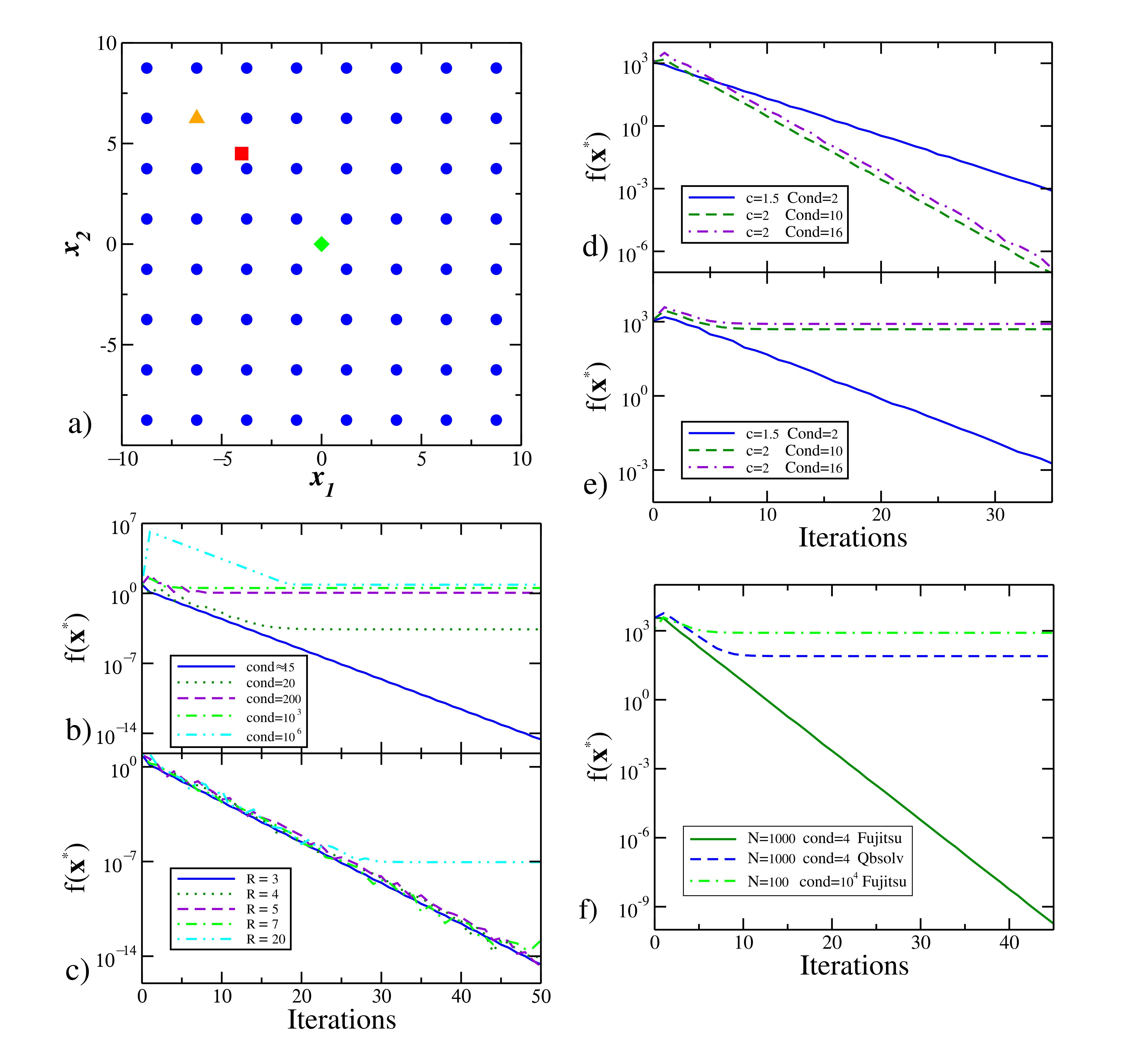}
 	\caption{Performance of the original algorithm for solving linear equation systems. (a) The $64=2^{3\times 2}$ vectors $\mathbf{x}(\mathbf{q})$ used to represent possible solutions, with $R=3$ e $N=2$. The green diamond corresponds with the initial guess $\mathbf{x}_0$, the red square is the exact solution, and the orange triangle is the vector $\mathbf{x}(\mathbf{q}^*)$ that minimizes the function $f(\mathbf{x})$ restricted to the possible 64 QUBO vectors, with $f\left(\mathbf{x(\mathbf{q}^*)}\right)=13/8$ and $\mathbf{q}^*=(0,0,1,1,1,0)$. In (b) and (c) we consider the iterative QUBO resolution of the linear system $\mathbf{A}\cdot \mathbf{x} = \mathbf{b}$, for matrices with different condition numbers and $N=2$. For each iteration, a vector $\mathbf{x}^*$ is obtained, and we plot $f(\mathbf{x}^*)$. We use $R=3$ in (b) and different values of $R$ until convergence is reached in (c). In both cases, we use $c=2$. The blue continue curve corresponds to the example in Eq. (\ref{LS1}). In (d) and (e) we consider linear systems with $N=100$ and matrix condition numbers $\mathrm{Cond}(\mathbf{A})=2, 10$, and $16$ (respectively, continue blue, green dashed, and dash-dot violet line). In Figure (d), we use the Fujitsu System as the QUBO solver. In Figure (e), we use Qbsolv in its standard configuration. In both cases, $R=3$, and the maximum time allowed per iteration is 30 seconds. In (f), we consider a linear system with $N=1000$ and a matrix condition number $\mathrm{Cond}(\mathbf{A})=4$. We use the Fujitsu System as the QUBO solver (solid green line) and the Qbsolv software in its standard configuration (blue dashed line). Additionally, we consider the case where $N=100$ and $\mathrm{Cond}(\mathbf{A})=10^4$ (green dashed-dot line), demonstrating that the method does not work well for square matrices with a large condition number. In all cases, $R=3$, and the maximum time allotted per iteration is 300 seconds.}
 	\label{FIG1}
\end{figure*}

The solution $\mathbf{x}^*=(-4,9/2)$ of the system minimizes the function 
\begin{equation}\label{Func}
	f(\mathbf{x})=\vert\vert \mathbf{A}\cdot\mathbf{x}-\mathbf{b} \vert\vert^2,
\end{equation}
with $f(\mathbf{x}^*)=0$, we choose $R=3$, $L=10$, and $\mathbf{x}_0=(0,0)$. The binary vectors $\mathbf{q}$ have $6$ components. In Figure \ref{FIG1}a, we depict the $2^6$ vectors to be analyzed. To construct the QUBO problem, we substitute Eq. (\ref{NormVectors}) into Eq. (\ref{Vectors}) and utilize the corresponding result in Eq. (\ref{Func}). It is not difficult to observe that the function (\ref{Func}) is redefined in the binary space of the 64 $\mathbf{q}$'s, and therefor we can construct a new  $N\times RN$ matrix $\mathbf{A}_{\mathbf{q}}$ and an $N$-vector $\mathbf{b}_{\mathbf{q}}$ satisfying

\begin{equation}\label{QUBOFunc}
	f(\mathbf{q})=\vert\vert \mathbf{A}_{\mathbf{q}}\cdot\mathbf{q}-\mathbf{b}_{\mathbf{q}} \vert\vert^2,
\end{equation}
where $\mathbf{A}_{\mathbf{q}}=\mathbf{A} \otimes (2^{0},2^{-1},2^{-2},\cdots,2^{1-R})$, with $\otimes$ the matrix kronecker product and $$\mathbf{b}_{\mathbf{q}}=\frac{1}{L}\left(\mathbf{b}+L\frac{\left(2^R-1\right)}{2^R}\mathbf{A}\cdot \mathbf{I}-\mathbf{A}\cdot \mathbf{x}_0\right).$$

In our particular case we have

\begin{equation}
	\mathbf{A}_{\mathbf{q}}=\begin{pmatrix}
		1 & 0.5 & 0.25 & 2 & 1 & 0.5\\ 	3 & 1.5 & 0.75 & 4 & 2 & 1\
	\end{pmatrix},\,\, \mathrm{and} \,\, \mathbf{b}_{\mathbf{q}}=\begin{pmatrix}
		3.125 \\ 6.75
	\end{pmatrix}.
\end{equation}

To construct the QUBO matrix used in eq. (\ref{QUBOFuncFinal}), we expand $f(\mathbf{q})=\left[(\mathbf{A}_{\mathbf{q}}\cdot\mathbf{q}-\mathbf{b}_{\mathbf{q}})\cdot(\mathbf{A}_{\mathbf{q}}\cdot\mathbf{q}-\mathbf{b}_{\mathbf{q}})\right]$. Neglecting the constant positive term $\mathbf{b}_{\mathbf{q}}\cdot\mathbf{b}_{\mathbf{q}}$, we obtain the symmetric QUBO matrix
\begin{equation}\label{QUBOMatrix}
\mathbf{Q}= \mathbf{A}_{\mathbf{q}}^{T}\cdot\mathbf{A}_{\mathbf{q}}-2*\mathrm{Diag}\left(\mathbf{A}_{\mathbf{q}}^{T}\cdot \mathbf{b}_{\mathbf{q}}\right),
\end{equation}
where $\mathrm{Diag}(\cdots)$ converts an $N$-vector into a diagonal $N\times N$ matrix. For our specific case, we have:

\begin{equation}
	\mathbf{Q}=\begin{pmatrix}
		-36.6 & 5 & 2.5 & 14 & 7 & 3.5\\ 	5 & -20.8 & 1.25 & 7 & 3.5 & 1.75 \\ 2.5 & 1.25 & -11.025 & 3.5 & 1.75 & 0.875 \\ 14 & 7 & 3.5 & -46.3 & 10 & 5 \\
		7 & 3.5 & 1.75 & 10 & -28.15 & 2.5 \\
		3.5 & 1.75 & 0.875 & 5 & 2.5 & -15.325
	\end{pmatrix}. \nonumber
\end{equation}

The binary vector $\mathbf{q}^*=(0,0,1,1,1,0)$ minimizes the function $f(\mathbf{q})$. In Figure \ref{FIG1}a, the orange triangle represents $x(\mathbf{q}^*)$, which minimizes the function in Eq. (\ref{QUBOFunc}). Note that in this case, the QUBO solution is not the closest point to the exact solution of the problem (the red square). However, for the procedure to work, it is necessary only that the orange configuration belongs to the same quadrant as the exact solution.

Once the vector $\mathbf{x}(\mathbf{q}^*)$ is found using a QUBO-solver, we repeat the process to find a better solution (closest to the exact solution $\mathbf{x}^*$) redefining $\mathbf{x}_0 \to \mathbf{x}(\mathbf{q}^*)$  and a new $L^*$, smaller than the previous $L$, in such way that the new $N$-cube contains a solution closer to the exact one.

For our concrete example ($RN=6$), verifying all the configurations and determining the best solution is easy. However, when $N$ is big, this procedure becomes intractable because the space of configurations is too large. A new search algorithm, different than the brute force approach, is necessary. There are different possibilities, such as simulated annealing algorithm \cite{Alkhamis1998}, Metaheuristic algorithms \cite{Dunning2018}, particular purpose quantum hardware such as quantum annealing machines  \cite{Hauke2020,Souza2021} and classical Ising machines \cite{Mohseni2022}. Hybrid procedures using quantum and classical computation are still possible \cite{Booth2020}.

\begin{figure}
	\centering
	\includegraphics[width=1\columnwidth]{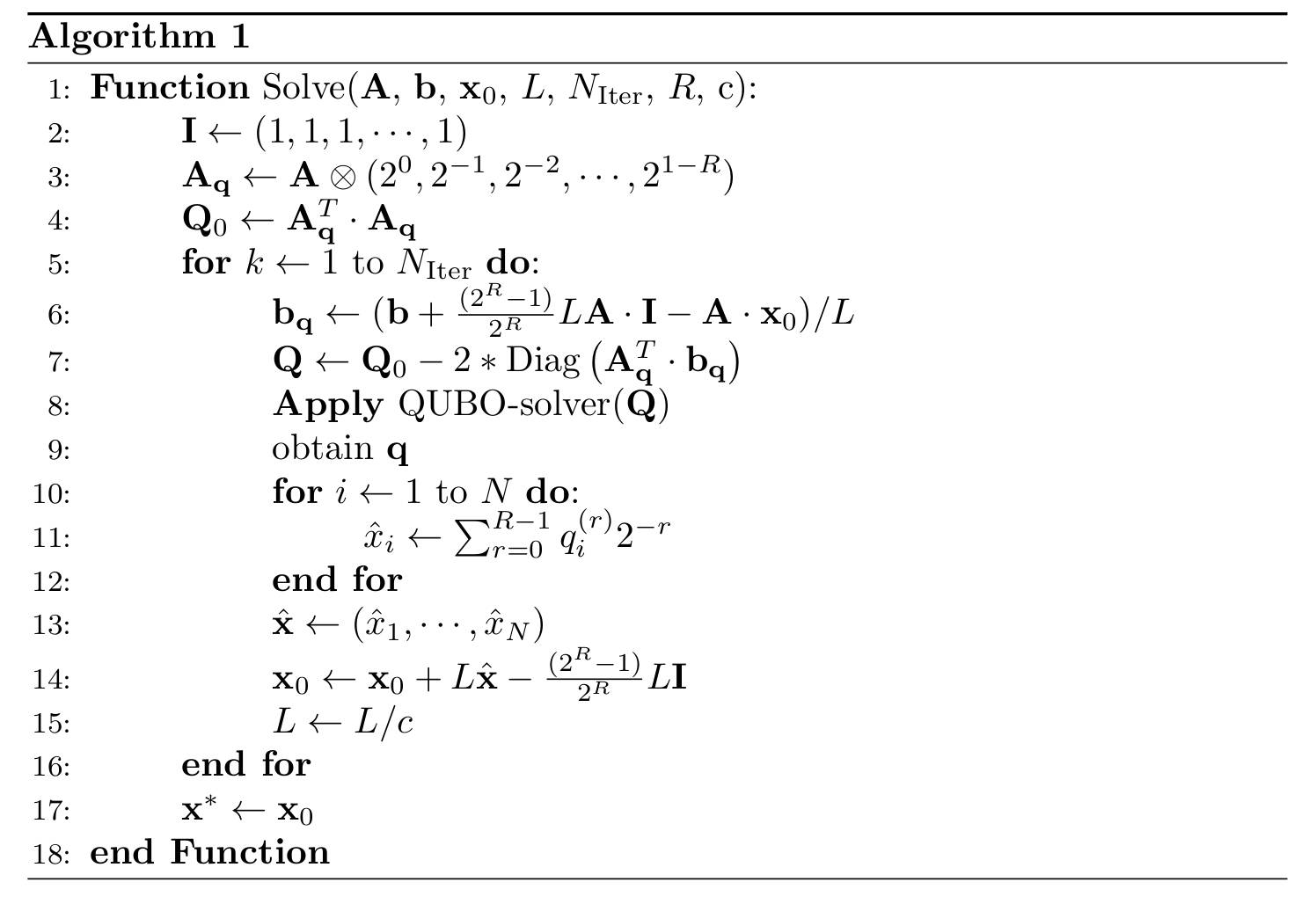}
	\caption{Preparation of the QUBO problem to solve a linear system of equations $\mathbf{A}\cdot \mathbf{x} = \mathbf{b}$, where $\mathbf{x}_0$ is the initial guess, $N_{\mathrm{Iter}}$ is the number of iterations used in the algorithm, $R$ is the bit approximation used for $\hat{x}_i$ and $c > 1$ is a constant.}
	\label{algo1}
\end{figure}

Other algorithms to tackle QUBO problems are mentioned in the review \cite{Kochenberger2014}. Once a QUBO-solver is chosen, we can use the iterative process to find the solution of the linear equations system. We implement this procedure in Algortithm 1, shown in Figure \ref{algo1}. 

\section{Methods}

After developing the appropriate mathematical tools, we implemented three methods in Python to solve the associated QUBO problem.

\begin{itemize}
\item \textit{Exhaustive Search (for small problems)}: When the number of variables (denoted by RN) is less than 20, we directly evaluate all possible QUBO configurations and select the one that yields the optimal solution. This approach is guaranteed to find the best solution but becomes computationally expensive for larger problems.
\item \textit{D-Wave QbSolv (deprecated)}: For larger problems, we employed the Qbsolv open-source software provided by D-Wave Systems (though it is currently deprecated). This Python library implements a simulated annealing algorithm, which we integrated into our code alongside Qbsolv.
\item \textit{Fujitsu Digital Annealer}: We additionally utilized the Fujitsu Digital Annealer system. We accessed the Fujitsu system through an Application Programming Interface (API) using Python's requests package. This allows our code to seamlessly interact with the Fujitsu system and submit QUBO problems for optimization.
\end{itemize}

The coefficients of the linear systems that we studied were randomly generated. After transforming these coefficients into a QUBO format, we converted them into JSON (JavaScript Object Notation) for efficient data exchange. The resulting JSON data was then sent to the Fujitsu system for optimization. Inquiries regarding the implementation details or the code itself can be directed to the authors.

\section{Results}

 In subsection \ref{Res1}, the performance of algorithm in Fig. \ref{algo1} applied to problems with a small condition number is shown. The algorithm works well in this case, but if we increase the condition number, convergence is only obtained by increasing the factor $R$ associated with the numerical binary approximation of the problem. Large condition numbers require larger $R$, and algorithm \ref{algo1} is no longer efficient.

In subsection \ref{Sec2}, the previous issue is addressed by determining the geometry of the hypersurfaces with $\mathbf{x}^T\cdot\left(\mathbf{A}^T\mathbf{A}\right)\cdot\mathbf{x}$ constant. We reformulate the QUBO problem considering this geometry and show that solving this problem is trivial even using $R=1$. A linear system consisting of $N = 5000$ equations with a condition number $10^6$ is solved, demonstrating the power of the method.

In subsection \ref{Sec3}, it is shown that partial knowledge of the geometry simplifies the QUBO approach. In particular, it is demonstrated that in a large problem with a condition number where the algorithm in Fig. \ref{algo1} fails, it is possible to decompose the original problem into many independent QUBO subproblems, each with a condition number amenable to being approached by the algorithm in Fig. \ref{algo1}. Such a decomposition is obtained by knowing the geometry only partially.

\subsection{Convergence of the conventional algorithm}\label{Res1}

The performance of Algorithm 1 strongly depends on the type of matrix $\mathbf{A}$ used in the problem, particularly on its condition number. The example described in Eq. (\ref{LS1}) has a condition number $\mathrm{Cond}(\mathbf{A}) \approx 15$. For this example, it is sufficient to use the parameters $R=3$ and $c=2$. As $\mathrm{Cond}(\mathbf{A})$ grows, the optimal QUBO configurations are further away from the exact solution of the problem, and it is possible that in the next iteration, the exact solution may fall outside of the $N$-cube, breaking convergence. This issue can be resolved by decreasing the parameter $c$, which increases the number of iterations needed to reach convergence.

Another option is to increase the factor $R$ of the algorithm, which increases the number of QUBO configurations. This, in turn, helps the optimal QUBO solution stay closer to the exact solution of the problem. However, increasing $R$ also enlarges the dimension of the QUBO problem to $RN \times RN$, thereby escalating the difficulty of the QUBO approach, at least in principle. In Figures \ref{FIG1}b-c, we illustrate these issues for the simpler case of $N=2$.

In Figures \ref{FIG1}d-e, we solve three different systems of linear equations with $N = 100$, $R=3$, and different $\mathrm{Cond}(\mathbf{A})<20$. The vector $\mathbf{b}$ associated with the problem was generated using random numbers between -200 and 200, and the matrix $\mathbf{A}$ was generated using random unitary transformations applied to appropriate diagonal matrices. Here, we compare the open-source heuristic algorithm Qbsolv in a classical simulation (which uses Tabu search and classical simulated annealing) and the Fujitsu system, which is a classical QUBO solver inspired by the quantum annealing approach. We observe that the Fujitsu system finds an adequate configuration in each iteration, reaching convergence when the process ends. For $N=100$, the Qbsolv software reaches convergence when $\mathrm{Cond}(\mathbf{A})=2$ and parameter $c=1.5$, showing that for $N=100$, it is advantageous to use the Fujitsu system.

Figure \ref{FIG1}f shows that for $N=1000$ and $\mathrm{Cond}(\mathbf{A})=4$, the method still works very well only for the fujitsu system. However, when $\mathrm{Cond}(\mathbf{A}) \gg 20$, the correspondence between optimal QUBO configurations that minimize Eq. (\ref{QUBOFunc}) and the closest configuration to the solution of $\mathbf{A}\cdot \mathbf{x} = \mathbf{b}$ is lost. We can choose a larger $R$ as shown in Figure \ref{FIG1}c, but for larger matrices with $\mathrm{Cond}(\mathbf{A}) \gg 20$, this procedure is not efficient.

The Fujitsu digital annealer enhances the well-known simulated annealing algorithm with other physics-inspired strategies that resemble quantum annealing procedures (see ref. \cite{Aramon2019}). In our case, involving large matrices, small binary approximations, and small condition numbers, the Fujitsu system seems to be very efficient at solving these types of problems. Large QUBO problems can be solved using the Fujitsu system (QUBO with dimensions up to $10^5$), which includes integration with the Azure system's blob storage to load even larger problems. However, even with an efficient QUBO solver like the Fujitsu system, in cases of large matrices with appreciable condition numbers and small binary approximations, algorithm 1 is not adequate for solving a linear system equation with an unique solution. For matrices with larger $\mathrm{Cond}(\mathbf{A})$, finding correspondence between QUBO configurations that minimize eq. (\ref{QUBOFunc}) and configurations sufficiently close to $\mathbf{x}^*$ depends on the initial guess $\mathbf{x}_0$. This property resembles the gradient descent algorithm used in minimization problems, where the convergence rate can heavily depend on the initial guess. This drawback is addressed in descent methods by considering the geometry of the problem and reformulating it into a more powerful conjugate gradient descent method. Next, we demonstrate that the geometry associated with the system of linear equations can improve convergence and break down a sizable original system $\mathbf{A}$ with an arbitrary $\mathrm{Cond}(\mathbf{A})$ number into smaller ones $\mathbf{A}_i$ with lower $\mathrm{Cond}(\mathbf{A}_i)$ that could be solved separately using algorithm 1.

\subsection{The rhombus geometry applied to the problem $\mathbf{A}\cdot \mathbf{x}=\mathbf{b}$}\label{Sec2}

\subsubsection{The geometry of the problem $\mathbf{A}\cdot \mathbf{x}=\mathbf{b}$}

\begin{figure*}
 	\centering
 	\includegraphics[width=2\columnwidth]{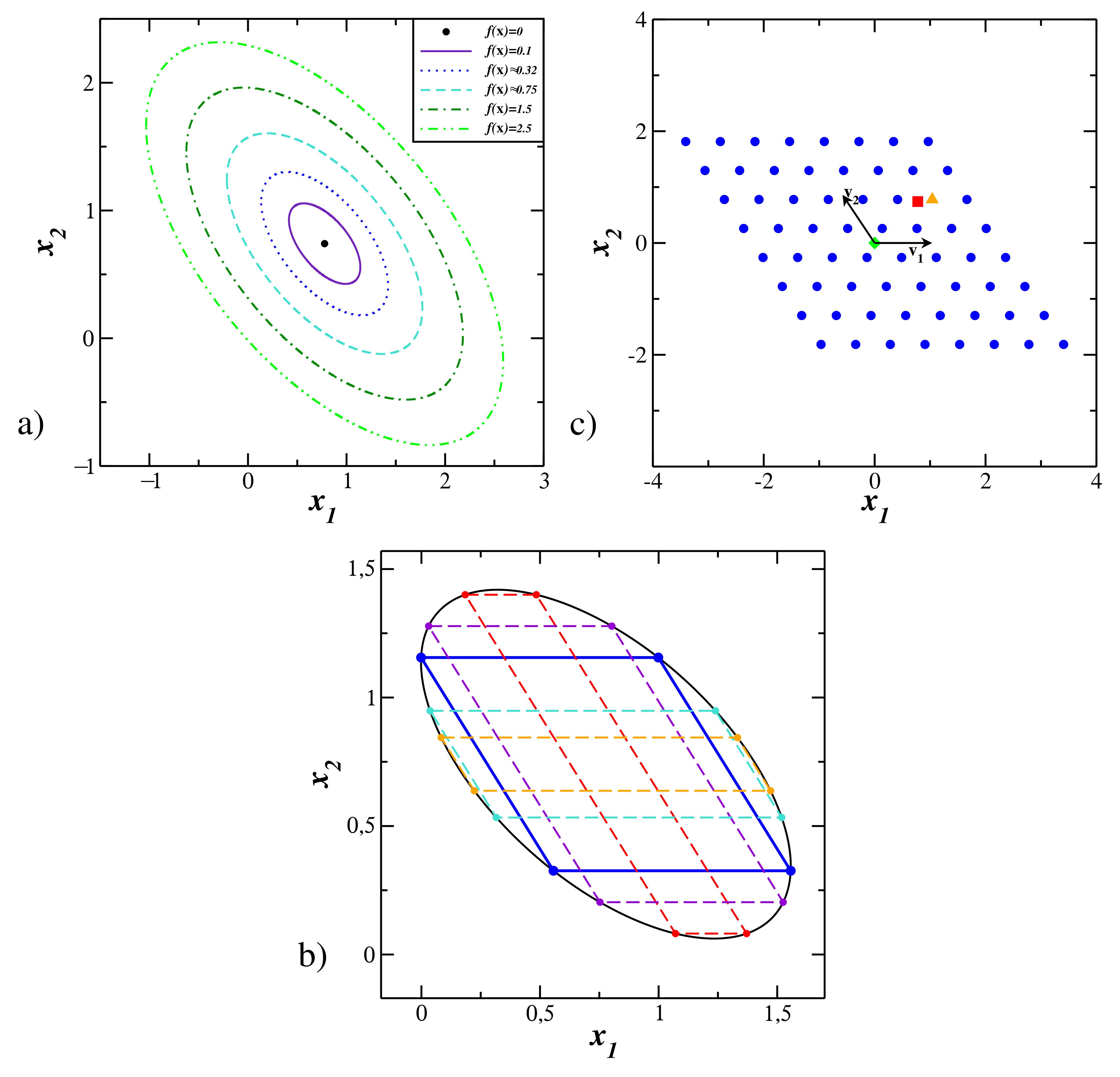}
 	\caption{Geometry of the matrix problem associated with the solution of a linear system. In (a), concentric ellipsoidal geometry in the inversion problem (\ref{Func}) for a particular case when $N=2$. $\mathbf{A}$ is invertible and the ellipsoids corresponds to the regions in $\mathbb{R}^N$ where $f(\mathbf{x})$ is constant. All the ellipsoids are concentric and contractible to the point $\mathbf{x}^*$ (black point), which is the unique point that satisfies $\mathbf{A}\cdot \mathbf{x}^*=\mathbf{b}$. In (b), family of parallelograms contained in a representative ellipse of the figure (a). The parallelograms have different side lengths and congruent angles. In the figure, we highlight the parallelogram with equal-length sides (rhombus) in blue, which would be the base of our method to solve the problem $\mathbf{A}\cdot \mathbf{x}=\mathbf{b}$. In (c), space configurations of the QUBO problem in the rhombus geometry (The green diamond corresponds with the initial guess $\mathbf{x}_0$, the red square is the exact solution, and the orange triangle is the point $\mathbf{x}(\mathbf{q}^*)$ that minimizes the function $f(\mathbf{x})$ between the lattice blue points). The unitary vectors $\mathbf{v}_1$ and $\mathbf{v}_2$ are the lattice vectors that define the rhombus geometry.}
 	\label{FIG2}
\end{figure*}

The entire discrete set of possible configurations defines the QUBO. Generally, there is little structure in this set. However, since the problem is written in the language of vector space, there is a robust mathematical structure that we can use to improve the performance of existing algorithms. It is not difficult to see that the subset of $\mathbb{R}^N$ where $f(\mathbf{x})$ (given by eq (\ref{Func}) with $\mathbf{A}$ invertible) is constant corresponds to ellipsoidal hyper-surfaces of dimension $N-1$. For $N=2$ see Figure \ref{FIG2}a.

All the ellipses in Figure \ref{FIG2}a are concentric and similar. Therefore, we can take a unique representative. Each ellipse contains a family of parallelograms with different sizes but congruent angles; see Figure \ref{FIG2}b. In Figure \ref{FIG1}a, the problem is formulated in a square lattice geometry. However, nothing prevents us from using another geometry, especially one adapted to the problem. We can choose a lattice with the parallelogram geometry. In particular, we choose the parallelogram with equal-length sides (rhombus). In Figure \ref{FIG2}c, we show how possible configurations are chosen using the rhombus geometry.  

The choice of this geometry brings advantages in the final algorithm efficiency, since we need only a few iterations with the rhombus geometry to obtain convergence to the solution. Given an initial guess $\mathbf{x}_0$, such point defines a rhombus. If  the solution $\mathbf{x}^*$ is also inside the same rhombus, then we can garantee that all subsequent steps will also be inside the same rombus as $\mathbf{x}^*$ (see proof in appendix \ref{RC}).  This property improves the convergence and will be called here as the rhombus convergence. 

 We emphasize that the square geometry used in previous works only coincides with the matrix inversion geometry when the matrix $\mathbf{A}$ is diagonal. For non-diagonal matrices in the square geometry, the closest point (in the conventional distance) to the exact solution $\mathbf{x}^*$ is not necessarily the point with the most negligible value of $f(\mathbf{x})$ between the finite QUBO vectors. In other words, the exact solution $\mathbf{x}^*$ would lay outside the region containing the QUBO configurations, breaking the convergence. We can avoid the lack of convergence by diminishing the parameter $c$ or increasing the number $R$ in the algorithm but with the consequence of increasing the number of iterations.

\subsubsection{$\mathbf{H}$-orthogonality}

The ellipsoid form in the matrix inversion problem is given by the symmetric matrix $\mathbf{H}=\mathbf{A}^{T}\mathbf{A}$, this becomes clear when we define the new function $f_0(\mathbf{x})=\Vert \mathbf{A}\cdot\mathbf{x}\Vert$, which define the same set of similar ellipsoids but centered in the zero vector. Particularly, the matrix $\mathbf{H}$ defins a different notion of orthogonality called in the review \cite{Shewchuk1994} as $\mathbf{H}$-orthogonality, where two vectors $\mathbf{v}_1$ and $\mathbf{v}_2$ in $\mathbb{R}^N$ are $\mathbf{H}$-orthogonal if they satisfy

\begin{equation}
\langle\mathbf{v}_1,\mathbf{v}_2\rangle_{\mathbf{H}}\equiv\mathbf{v}_1\cdot\left(\mathbf{A}^{T}\mathbf{A}\cdot \mathbf{v}_2\right)=0.
\end{equation}

Given the $N$'s canonical vectors $\mathbf{u}_k$, with the $k$-th coordinate equal to one and all others equal to zero, we can construct from them $N$ $\mathbf{H}$-orthogonal vectors $\mathbf{v}_k$ associated with each $\mathbf{u}_k$ using a generalized Gram-Schmidt $\mathbf{H}$-orthogonalization. The method selects the first vector as $\mathbf{v}_1=\mathbf{u}_1$. The vector $\mathbf{v}_{m}$ is constructed as  

\begin{equation}\label{GramVect}
	\mathbf{v}_{m}=\mathbf{u}_{m}+\sum_{k=1}^{m-1}\beta_{mk}\mathbf{v}_k.
\end{equation}

The coefficients $\beta_{mk}$ are determined using the $\mathbf{H}$-orthogonality property $\langle\mathbf{v}_m,\mathbf{v}_k\rangle_{\mathbf{H}}=0$. Explicitly

\begin{equation}
	\beta_{mk}=-\frac{\langle\mathbf{v}_k,\mathbf{u}_m\rangle_{\mathbf{H}}}{\langle\mathbf{v}_k,\mathbf{v}_k\rangle_{\mathbf{H}}},
\end{equation}
\begin{figure}
	\centering
	\includegraphics[width=1\columnwidth]{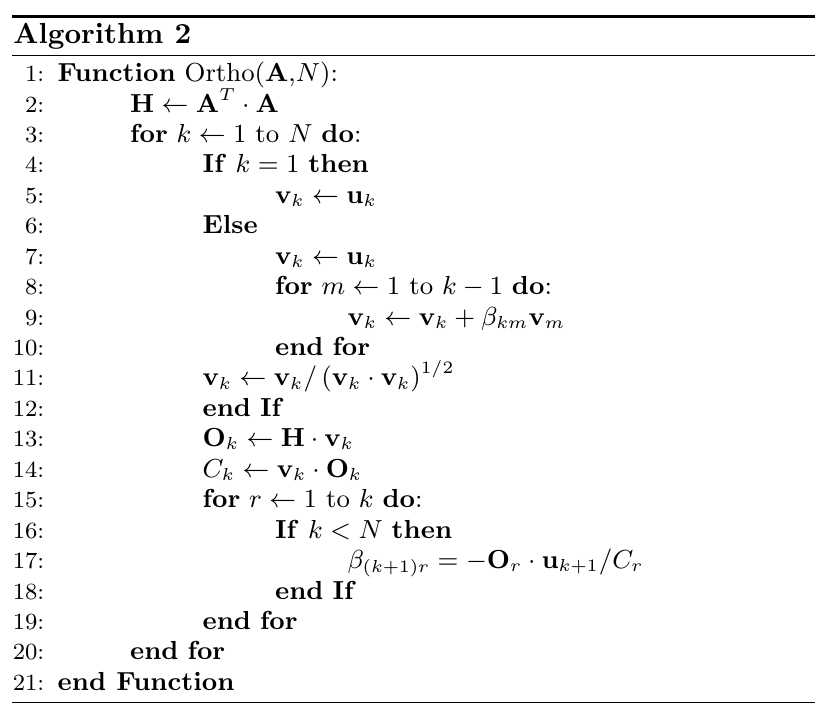}
	\caption{Gram-Schmidt procedure for the calculus of the $N$’s $\mathbf{H}$-orthogonal vectors $(\mathbf{v}_1,\cdots,\mathbf{v}_N)$}
	\label{algo2}
\end{figure}
implemented this procedure in Algortithm 2, shown in Figure \ref{algo2}. The calculated non-orthogonal unitary vectors (in the standard scalar product) $\mathbf{v}_k$ define the rhombus geometry previously described. In Appendix \ref{RC2}, we improve the algorithm described above.

\subsubsection{The modified search region}

Considering the intrinsic rhombus geometry, the iterative algorithm converges exponentially fast in the number of iterations and is sufficient to use $R=1$. The QUBO configurations in (\ref{Vectors}) around a certain guess $\mathbf{x}_0$ can be rewritten as

\begin{equation}\label{Vectors2}
	\mathbf{x}=\mathbf{x}_0+L\sum_{i=1}^N\left(\hat{x}_i-1/2\right)\mathbf{u}_i,
\end{equation}
where $\{\mathbf{u}_i\}$ is the canonical base. Therefore, we modified Algorithm 1 changing $\mathbf{u}_i \to \mathbf{v}_i$ and $\hat{x}_i \to q_i$.  We have $q_i \in \{0,1\}$. The $2^N$ QUBO configurations are the vertices of a $N$-rhombus and are associated with all the possible binary vectors $\mathbf{q}=(q_1,\cdots,q_N)$. We can substitute this modifications in the function $f(\mathbf{x})$ and calculate $\mathbf{A}_{\mathbf{q}}$ and $\mathbf{b}_{\mathbf{q}}$. Considering the vectors $\mathbf{v}_i$ as the $i$th row of a matrix $\mathbf{V}$ (this is $V_{ij}=\mathbf{v}_i\cdot \mathbf{u}_j$) is not difficult to see that

\begin{equation}\label{QUBOA}
	\mathbf{A}_{\mathbf{q}}=\mathbf{A}\mathbf{V}^{T}
\end{equation}
and
\begin{equation}
\mathbf{b}_{\mathbf{q}}=(\mathbf{b}+\frac{L}{2}\mathbf{A}_{\mathbf{q}}\cdot \mathbf{I}-\mathbf{A}\cdot \mathbf{x}_0)/L.
\end{equation}

\begin{figure}
	\centering
	\includegraphics[width=1\columnwidth]{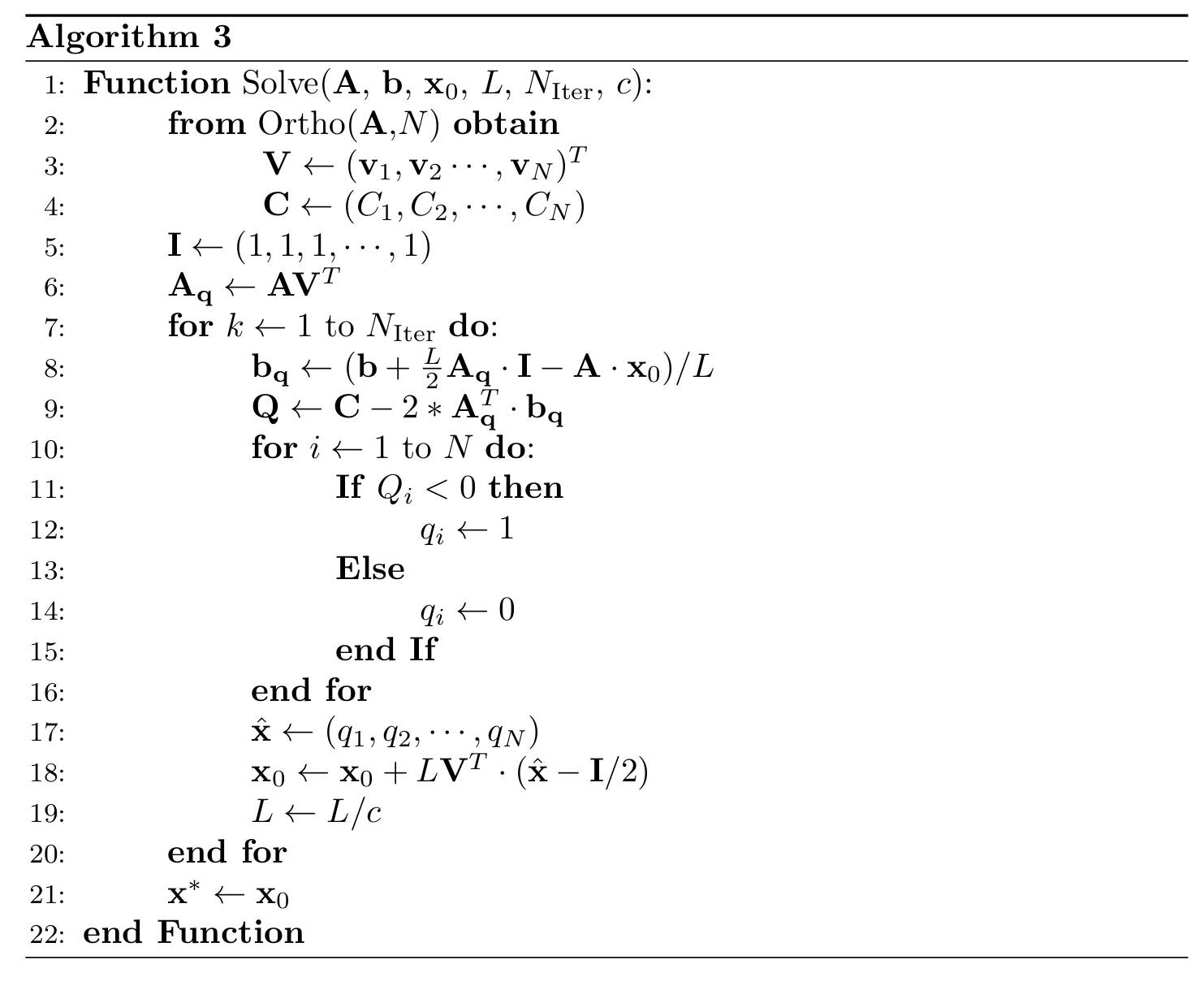}
	\caption{Modified iterative algorithm using the rhombus geometry. The $C_k$ numbers are calculated in Algorithm 2.}
	\label{algo3}
\end{figure}

From equation (\ref{QUBOMatrix}) and the $\mathbf{H}$-orthogonality of the vectors $\mathbf{v}_i$ (matrix rows of $\mathbf{V}$), it is possible to see that the QUBO matrix $\mathbf{Q}$ constructed from (\ref{QUBOA}) is always diagonal. The QUBO solution is trivial (this means that there are no necessary heuristic algorithms or quantum computers to solve the QUBO problem). The modified iterative process is shown in Algorithm 3, shown in Figure \ref{algo3}

\subsubsection{Implementation of the  algorithm}

\begin{figure*}
 	\centering
 	\includegraphics[width=2\columnwidth]{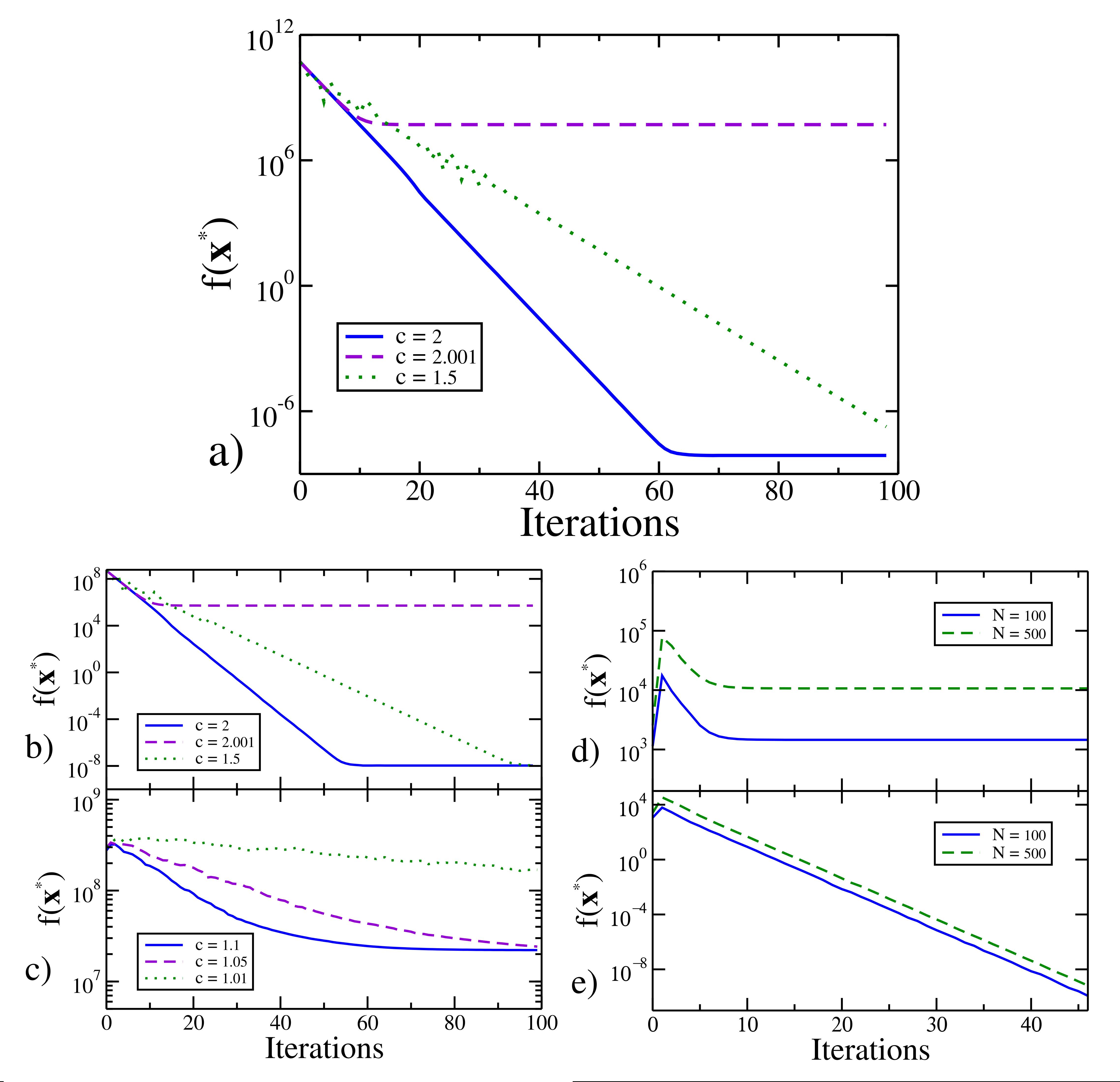}
 	\caption{Performance of the new method for solving linear systems. In (a), we shown the iterative QUBO modified algorithm applied to a linear system with 5000 variables and 5000 equations and $\mathrm{Cond}(\mathbf{A}) \approx 10^6$. Note the fast convergence rate to the exact solution in a few iterations (cases $c=2$ and $c=1.5$). All the $2.5\times 10^7$ matrix coefficients of $\mathbf{A}$ and the 5000 vector coefficient of $\mathbf{b}$ were generated using random numbers between 0 and 200. In the modified algorithm we use $L=61000$ and initial guess $\mathbf{x}_0=(0,0,\cdots,0)$. Considering $\mathbf{x}_{\mathrm{Inv}}=\mathbf{A}^{-1}\cdot \mathbf{b}$ as the solution obtained by classical inversion algorithms, we have $f(\mathbf{x}_{\mathrm{Inv}})\approx 7.07\times 10^{-7}$. We obtain $f(\mathbf{x}^*)\approx 7.08\times 10^{-9}$ with our modified QUBO algorithm. For $c>2$, the convergence is drastically destroyed. In (b) and (c), we have the comparison between the iterative algorithms 3 (Figure (b)) and algorithm 1 (Figure (c)) for a matrix with $N=500$, $\mathrm{Cond}(\mathbf{A})\approx 10^6$ and the same initial $L$. The modified algorithm performs substantially better than the original (see Figure (b)). The function $f(\mathbf{x}^*)$ for the vector $\mathbf{x}^*$ obtained in the final iteration is very close to zero). Algorithm 1 (Figure (c)) using Qbsolv as QUBO-solver in the standard configuration does not show convergence. The Fujitsu system present the same behavior (We tested only 20 iterations, and the figure is not shown). In the two cases, the initial guess $\mathbf{x}_0$ is the zero vector. In (d) and (e), we have the iterative algorithms 1 and 5 applied to two linear systems with dimensions $100\times 100$ and $500\times 500$. All the matrix coefficients of $\mathbf{A}$ and the vector coefficients of $\mathbf{b}$ were generated using random numbers between $-200$ and $200$. We use $L=100$ and an initial guess $\mathbf{x}_0=(0,0,\cdots,0)$. In (d), we use the square geometry and algorithm 1. In Figure (e), we decompose the original matrix into 10 sub-problems with dimensions $10\times 10$ and 10 sub-problems with dimensions $50\times 50$. We only obtain exact convergence to the solution using the block decomposition shown in algorithm 5. In all cases, we use $R=3$ and $c=2$.}
 	\label{FIG3}
\end{figure*}

The algorithm 3 works whenever the rhombus that contains the QUBO configurations also includes the exact solution $\mathbf{x}^*$. This is guaranteed when $L$ is sufficiently large (in particular when $L>L_0$, where $L_0$ is the ``critical” value parameter to obtain convergence). In Figure \ref{FIG3}a, we show the algorithm performance for a particular dense matrix with dimension $5000\times5000$. The initial guess is the $N$-dimensional zero vector ($N=5000$). Notice the dependence with the parameter value $c$; the critical value is $c=2$, and there is no convergence for $c>2$. To compare with the original algorithm, we study the case with $N=500$, corresponding to a QUBO problem with 1500 variables ($N = 500$ and $R = 3$). Figures \ref{FIG3}b-c compares the two different approaches. The original Algorithm 1 in Figure \ref{FIG3}c has poor efficiency compared to the modified Algorithm 3 shown in Figure \ref{FIG3}b. 

In the last section of this work, we show that a partial knowledge of the conjugated vectors $\mathbf{v}_i$ also simplifies the original QUBO problem considerably. 

\subsection{Solving Large systems of equations using binary optimization}\label{Sec3}

\subsubsection{ Decomposing QUBO matrices in smaller sub-problems.}

In the previous section, we show that the knowledge of the conjugated vectors that generate the rhombus geometry simplifies the QUBO resolution and improves the convergence rate to the exact solution. However, the calculus of these vectors in Algorithm 2 has approximately $O(N^3)$ steps. A faster algorithm would be desirable.

Another interesting possibility is to use the notion of $\mathbf{H}$-orthogonality to construct a different set of  $N$ vectors $\mathbf{v}_i$ grouped in $m$ different subsets in such a way that vectors in different subsets are $\mathbf{H}$-orthogonal. In this last section, we show that such construction decomposes the original QUBO matrix in a Block diagonal form, and we can use a modified version of Algorithm 3. We can tackle each block independently for some QUBO-solver, and after joining the independent results, we obtain the total solution. There are a total of $B_N$ possible decompositions,
where $B_N$ is the number of possible partitions of a set
with N elements (Bell numbers).

Techniques of decomposition in sub-problems are standard in the search process in some QUBO-solvers. We mention the QUBO-solver Qbsolv, a heuristic hybrid algorithm that decomposes the original problem into many QUBO sub-problems that can be approached using classical Ising or Quantum QUBO-solvers. The solution of each sub-problem is projected in the actual space to infer better initial guesses in the classical heuristic algorithm (Tabu search); see \cite{Booth2017} for details. Our algorithm decomposes the original QUBO problem associated with $\mathbf{A}\cdot\mathbf{x} = \mathbf{b}$ in many independent QUBO sub-problems. We obtain the optimal solution directly from the particular sub-solutions of each QUBO sub-problem.

To see how the decomposition method works, we use the generalized Gram-Schmidt orthogonalization only between different groups of vectors. We choose $m$ positive numbers $a_i$, satisfying $N=a_1+a_2+\cdots+a_m$. First, call

\begin{equation}
\mathbf{v}_i^{(1)}=\mathbf{u}_i,\,\,\,\,\,\,\,\,\mathrm{If}\,\,\,\,\,\,\,\, i \in \{1,\cdots,a_1\}.
\end{equation}  

For the other vectors, we use

\begin{equation}\label{BlockV}
\mathbf{v}_j^{(1)}=\mathbf{u}_j+\sum_{k=1}^{a_1}\beta_{jk}\mathbf{v}_k^{(1)},\,\,\,\,\,\,\,\,\mathrm{If}\,\,\,\,\,\,\,\, j \in \{a_1+1,\cdots,N\}.
\end{equation} 

We also demand that the first group of $a_1$ vectors be $\mathbf{H}$-orthogonal to the second group of $N-a_1$ vectors, this is for $j \in \{a_1+1,\cdots,N\}$:

\begin{align}
\langle \mathbf{v}_k^{(1)} ,\mathbf{v}_j^{(1)}\rangle_{\mathbf{H}}=0,\,\,\,\mathrm{If}\,\,\,\, & k \in \{1,\cdots,a_1\} \,\,\, \mathrm{and} \,\,\, \nonumber \\ &j \in \{a_1+1,\cdots,N\}.
\end{align}

\begin{figure}
	\centering
	\includegraphics[width=1\columnwidth]{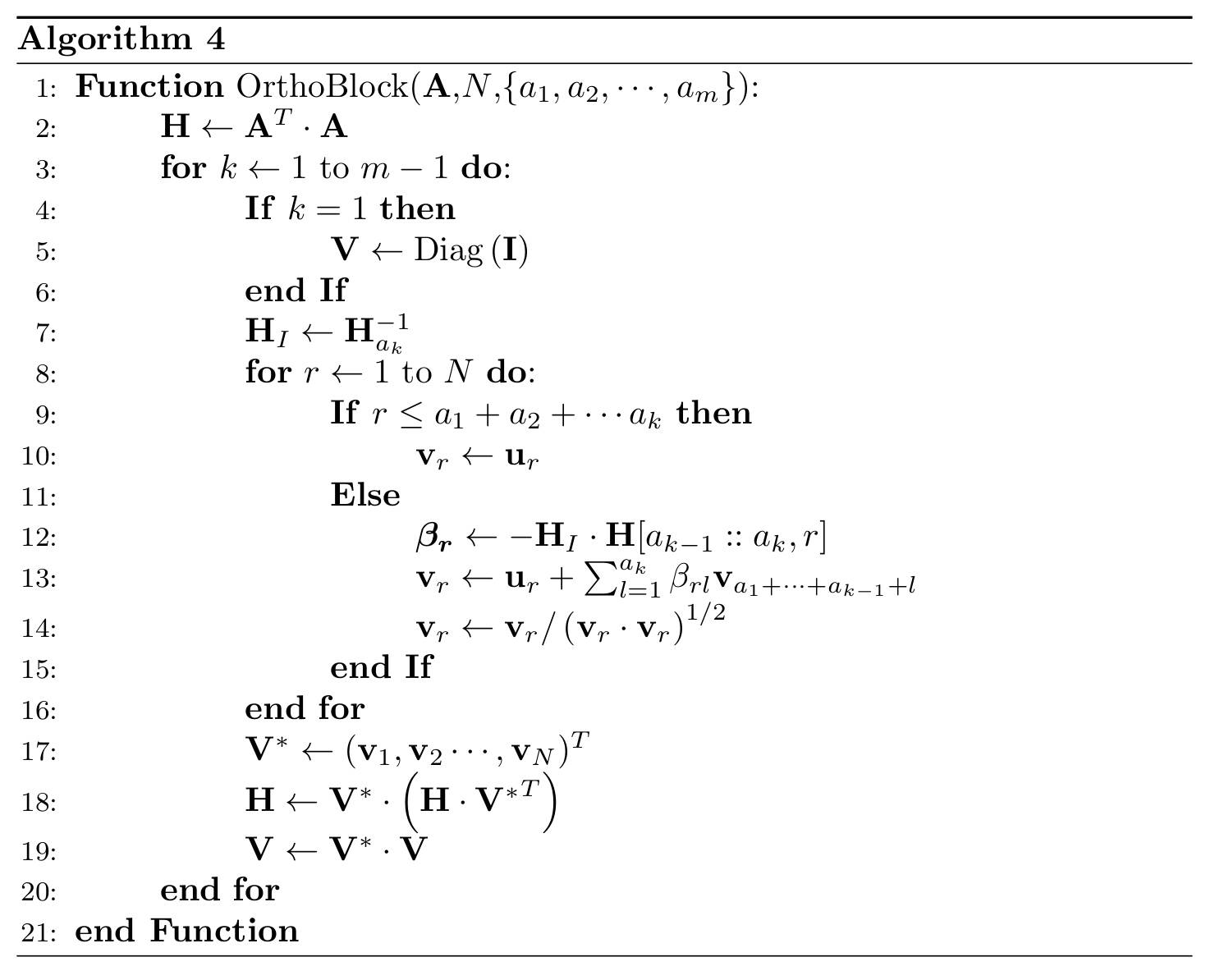}
	\caption{Block diagonal transformation of the matrix $\mathbf{H}$ associated with the composition $(a_1,a_2,\cdots,a_m)$ from the partial $\mathbf{H}$-orthogonalization process described in the construction of equation (\ref{OrthoBlock}). In the pseudo-code the notation is $\bm{\beta}_r=(\beta_{r1},\beta_{r2},\cdots,\beta_{ra_k})$ and $\mathbf{H}[a_{k-1}::a_{k},r]$ correspond with the $r$ sub-column of $\mathbf{H}$ beginning in the row component $a_1+\cdots+a_{k-1}+1$ and finishing in $a_1+\cdots+a_{k-1}+a_k$.}
	\label{algo4}
\end{figure}

This last condition determines all the coefficients $\beta_{jk}$ for each $j$, solving a linear system of dimension $a_1\times a_1$. For fixed $j$ and defining $\bm{\mathbf{\beta}}_{j}=(\beta_{j1},\beta_{j2},\cdots,\beta_{ja_1})$, the linear system to solve is

\begin{equation}
\bm{\mathbf{\beta}}_{j}=-\mathbf{H}_{a_1}^{-1}\cdot \mathbf{h}_{j},
\end{equation}
where $\mathbf{H}_{a_1}$ is the corresponding sub-matrix of $\mathbf{H}$ with their first $a_1\times a_1$ block sub-matrix, and $\mathbf{h}_j$ are the first $a_1$'s coefficients of the $j$-column of $\mathbf{H}$. With the coefficients $\bm{\beta}_j$ we can calculate $\mathbf{v}_j^{(1)}$ and normalize it. Grouping all these vectors as the rows of the matrix $\mathbf{V}_{(1)}$, it is possible to verify that

\begin{equation}
\mathbf{V}_{(1)}\cdot \mathbf{H} \cdot \mathbf{V}_{(1)}^T= \mathbf{H}_{a_1}\oplus\mathbf{H}^{(1)},
\end{equation}
where $\mathbf{H}^{(1)}$ is a $(N-a_1)\times(N-a_1)$ matrix. We can put $\mathbf{H}^{(1)}$ in a two-block diagonal form by the same process, where one block has dimension $a_2\times a_2$ and the second block has dimension $(N-a_1-a_2)\times(N-a_1-a_2)$. That is: 

\begin{equation}\label{IndexF}
\mathbf{v}_i^{(2)}=\mathbf{u}_i,\,\,\,\,\,\,\,\,\mathrm{If}\,\,\,\,\,\,\,\, i \in \{0,1,\cdots,a_1+a_2\},
\end{equation}
and
\begin{equation}\label{BlockV1}
\mathbf{v}_j^{(2)}=\mathbf{v}_j^{(1)}+\sum_{k=a_1+1}^{a_1+a_2}\beta_{jk}^{(1)}\mathbf{v}_k^{(2)},\,\,\,\,\,\mathrm{If}\,\,\,\,\, j \in \{a_1+a_2+1,\cdots,N\}.
\end{equation}
To determine the new set of $\beta$ coefficients, we use

\begin{equation}
\bm{\mathbf{\beta}}_{j}^{(1)}=-{\mathbf{H}_{a_2}^{-1}}\cdot \mathbf{h}_{j}^{(1)},
\end{equation}
where $\bm{\mathbf{\beta}}_{j}^{(1)}=(\beta_{j,a_1+1}^{(1)},\beta_{j,a_1+2}^{(1)},\cdots,\beta_{j,a_1+a_2}^{(1)})$, $\mathbf{H}_{a_2}$ is the first $a_2\times a_2$ block diagonal matrix of $\mathbf{H}^{(1)}$ and $\mathbf{h}_{j}^{(1)}$ are the first $a_2$'s coefficients of the $j$-column of $\mathbf{H}^{(1)}$.  Repeating the previous procedure, we obtain a new matrix $\mathbf{V}_{(2)}$, which has the property 
\begin{equation}
\mathbf{V}_{(2)}\cdot\left(\mathbf{V}_{(1)}\cdot \mathbf{H} \cdot \mathbf{V}_{(1)}^T\right)\cdot \mathbf{V}_{(2)}^T= \mathbf{H}_{a_1}\oplus\mathbf{H}_{a_2}\oplus\mathbf{H}^{(2)}.
\end{equation}
 Repeating the same process another $(m-3)$ times and defining
\begin{equation}
\mathbf{V} \equiv \mathbf{V}_{m-1}\cdot\mathbf{V}_{m-2}\cdots\mathbf{V}_{2}\cdot\mathbf{V}_{1}
\end{equation}
and $\mathbf{H}^{(m-1)}\equiv \mathbf{H}_{a_m}$, we obtain

\begin{equation}\label{OrthoBlock}
\mathbf{V}\cdot\mathbf{H}\cdot\mathbf{V}^{T}=\mathbf{H}_{a_1}\oplus\mathbf{H}_{a_2}\oplus\cdots\oplus\mathbf{H}_{a_m}
\end{equation}
We use the notation $\mathbf{H}_{a_k}$ to reinforce that this is a $a_k\times a_k$ matrix. We implemented this procedure in Algortithm 4, shown in Figure \ref{algo4}.

To effectively decompose a large matrix $\mathbf{A}$ with an arbitrary condition number $\mathrm{Cond}(\mathbf{A})$ into $m$ sub-problems ${\mathbf{H}_{a_1},\cdots,\mathbf{H}_{a_m}}$, each tractable with algorithm 1, we need to choose adequate submatrices of $\mathbf{H}=\mathbf{A}^T\cdot\mathbf{A}$ with $\mathrm{Cond}(\mathbf{H}_{a_i})<\sqrt{15}$. This is always possible for large matrices $\mathbf{H}$ using the following procedure: To construct $\mathbf{H}_{a_1}$, first test all the $2\times 2$ submatrices of $\mathbf{H}$ and choose the one with the minimal condition number. Next, test all the $N-2$ remaining indices to construct a $3\times 3$ matrix with the previous matrix and choose the one with the minimal condition number, repeating the procedure until reaching the desired dimension $a_1\times a_1$ to obtain $\mathbf{H}_{a_1}$. Then apply the modified orthogonalization procedure explained above to obtain a new matrix $\mathbf{H}^{(1)}$ with dimension $(N-a_1)\times(N-a_1)$, and with this matrix construct $\mathbf{H}_{a_2}$ in the same way, repeating the procedure until reaching $\mathbf{H}_{a_m}$. Evidently, the indices of the submatrices are not ordered, but the generalization is straightforward: with each matrix $\mathbf{H}_{a_i}$, there is associated a set of indices $\mathcal{A}_i \subseteq \{1,\cdots,N\}$, with $\mathcal{A}_i \cap \mathcal{A}_j=\emptyset$ if $i \neq j$, and in eq. (\ref{IndexF}), the substitution is $i \in \mathcal{A}_1 \cup \mathcal{A}_2$. It is not difficult to show that there exists a permutation $\sigma$ of the matrix indices such that the row-column permutation $\mathbf{P}(\sigma)\cdot\left(\mathbf{V}\cdot\mathbf{H}\cdot\mathbf{V}^T\right)\cdot\mathbf{P}(\sigma)^{-1}$ is put into explicit block-diagonal form. This remark is important because we need to manipulate each block independently as we going to show in the next section.

\subsubsection{Implementation of the  algorithm}
\begin{figure}
	\centering
	\includegraphics[width=1\columnwidth]{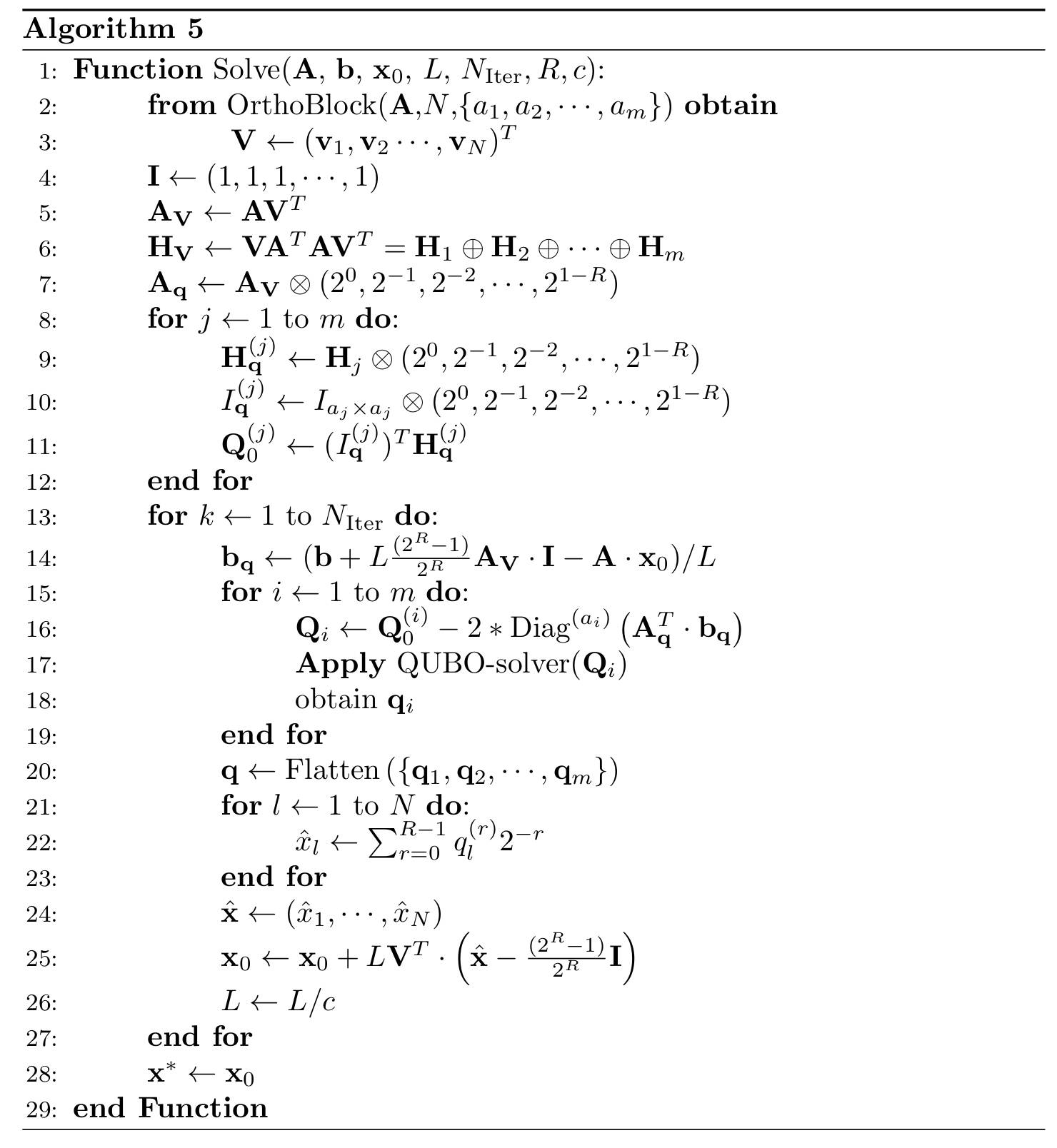}
	\caption{Modified iterative algorithm using the block diagonal decomposition of $\mathbf{H}$.  Here, $\mathrm{Diag}^{(a_i)}\left(\mathbf{A}_{\mathbf{q}}^{T}\cdot \mathbf{b}_{\mathbf{q}}\right)$ take the components of the vector $\left(\mathbf{A}_{\mathbf{q}}^{T}\cdot \mathbf{b}_{\mathbf{q}}\right)$ from the coordinate $R\times(a_1+\cdots+a_{i-1})+1$ until the coordinate $R\times(a_1+\cdots+a_{i})$ and builds a diagonal $Ra_i\times Ra_i$ matrix.}
	\label{algo5}
\end{figure}

 Suppose that the matrix $\mathbf{H}$ is placed in block diagonal form with each block $\mathbf{H}_i$ having $\mathrm{Cond}(\mathbf{H}_i)<\sqrt{15}$, as explained above. The procedure that decomposes and solves a QUBO problem is shown in algorithm 5, Figure \ref{algo5}. For each matrix $\mathbf{H}_i$ in each iterative step, the Fujitsu QUBO-solver system is used. In Figures \ref{FIG3}d-e, we present the resolution of two matrices with sizes $100\times 100$ and $500\times 500$ using block decomposition into ten $10\times 10$ subproblems and ten $50\times 50$ subproblems, respectively. The condition number of both matrices is respectively $\mathrm{Cond}(\mathbf{A})=1.3\times 10^5$ and $\mathrm{Cond}(\mathbf{A})=8.6\times 10^5$. Note that our method, unlike the original algorithm 1, works for arbitrary matrices and is not restricted to matrices with small condition numbers (the two matrices were generated by choosing random integers in the interval $[-200,200]$).

\section{Discussion}\label{Sec4}

It has recently been conjectured that the use of quantum technologies would improve the learning process in machine learning models. In the standard quantum circuit paradigm, many proposals and generalizations exist, promising better performance with the advent of quantum computers. Machine learning formulations as QUBO problems are also another possible strategy that can be improved with the development of quantum annealing hardware. In such cases, the approach of linear algebra problems as QUBO is of general interest because linear algebra is one of the natural languages in which machine learning is written. In this work, we proposed a new method to solve a system of linear equations using binary optimizers. Our approach guarantees that the optimal configuration is the closest to the exact solution. Additionally, we demonstrated that partial knowledge of the problem's geometry allows decomposition into a series of independent sub-problems that can be solved using conventional QUBO solvers. The solution to each sub-problem is then aggregated, enabling rapid determination of an optimal solution. We show that the original formulation as QUBO is efficient only when the condition number of the associated matrix $\mathbf{A}$ is small (with $\mathbf{A}$ being a square matrix). Our procedure is applicable in principle to matrices with arbitrary condition numbers where the error associated with the multiplication operations is controlled. Therefore, our method is not restricted to matrices with condition numbers close to one.

However, identifying the vectors that determine the sub-problem decomposition incurs computational costs that influence the overall performance of the algorithm. Nevertheless, two factors could lead to significant improvements: better methods for identifying the vectors associated with the geometry and faster QUBO solvers. In our study, when using a QUBO solver such as the Fujitsu digital annealer, we focus on finding elite QUBO solutions. This is because, when the condition number is small, we are guaranteed that the associated configuration is very close to the solution of the problem. Finding elite solutions to QUBO problems is very costly for large problems due to their NP-hardness. However, the only criterion for obtaining convergence in Algorithm 1 is to get a configuration in the same quadrant that contains the solution of the linear system of equations. The number of configurations in each quadrant (there are $2^N$ quadrants) is $2^{RN}/2^N$. For large $N$, this results in a large number of configurations. Therefore, focusing on developing new methods to find configurations in the same quadrant as the solution would be an interesting strategy to overcome the NP-hardness of finding the best QUBO solution. In any case, quantum computing, or quantum inspired classical computation could be fundamental tools for developing better approaches that can be integrated with the procedures presented here. We intend to explore these interesting questions in subsequent studies, and we hope that the methods presented here can contribute to the discovery of better and more efficient procedures for solving extensive linear systems of equations.

\begin{acknowledgements}
This work was supported by the Brazilian National Institute of Science and Technology for Quantum Information (INCT-IQ) Grant No. 465 469/2 014-0, the Coordenação de Aperfeiçoamento de Pessoal de Nível Superior—Brasil (CAPES)—Finance Code 001, Conselho Nacional de
Desenvolvimento Científico e Tecnológico (CNPq) and PETROBRAS: Projects 2017/00 486-1, 2018/00 233-9, and 2019/00 062-2. AMS acknowledges support from FAPERJ (Grant No. 203.166/2 017). ISO acknowledges FAPERJ (Grant No. 202.518/2 019).
\end{acknowledgements}

\bibliographystyle{ieeetr}
\bibliography{RefsArticle}

\appendix

\section{The rhombus convergence}\label{RC}

\begin{figure}[h]
	\centering
	\includegraphics[width=1\columnwidth]{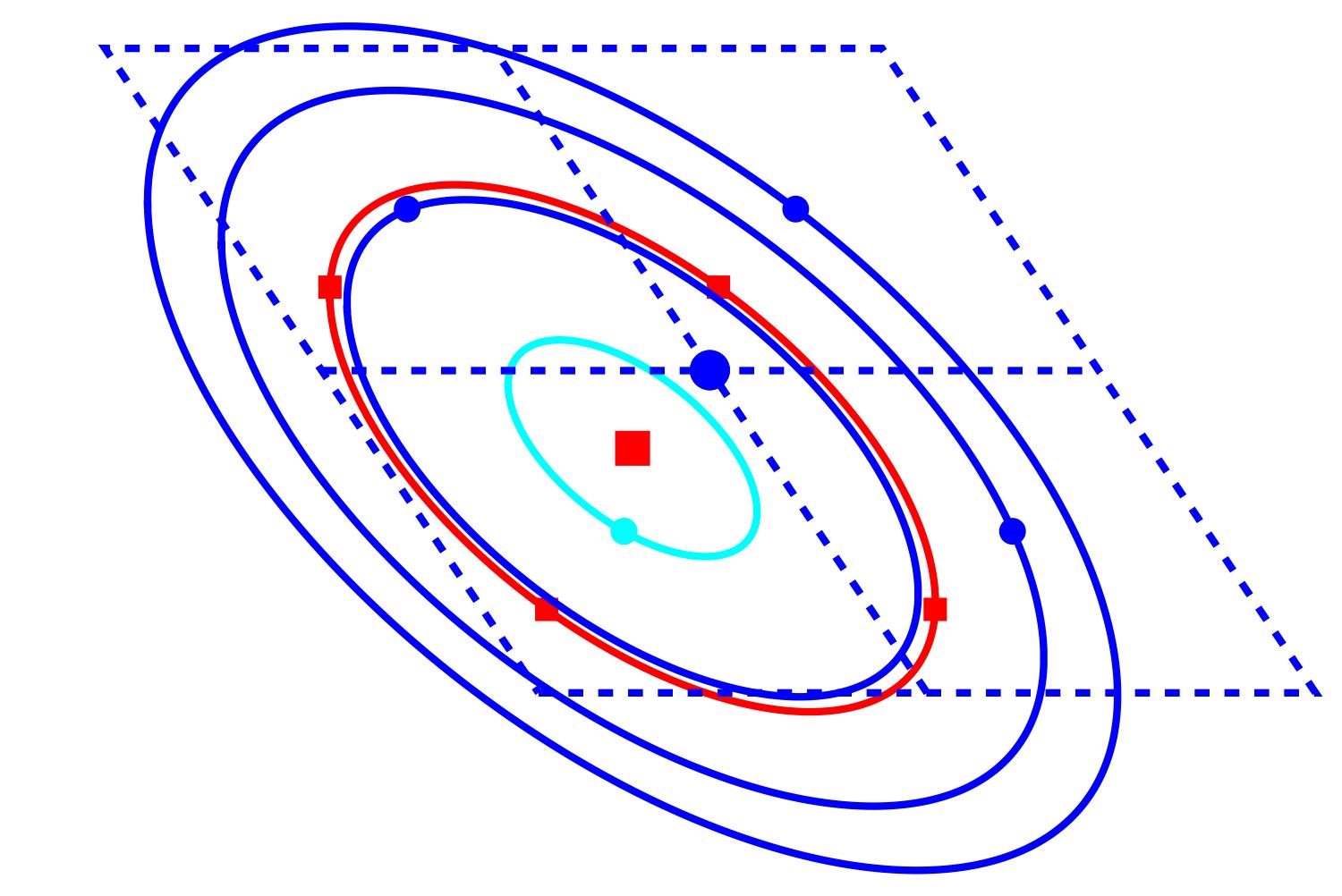}
	\caption{The figure shows the rhombus defined by the geometry of the problem; the cyan and blue points are the QUBO configurations $\mathbf{y}_k$  around the big blue point $\mathbf{x}_0$ in the center of the rhombus. There are $2^N$ points (in the figure, $N=2$), and with each point, an associated sub-rhombus. The big red square $\mathbf{x}^*$ satisfy $f(\mathbf{x}^*)=0$ and $\mathbf{x}^*=\mathbf{x}_0+\mathbf{t}$, with $\mathbf{t}$ the difference vector. In the figure, the red tiny squares satisfy $\mathbf{x}_k=\mathbf{y}_k+\mathbf{t}$ and $f(\mathbf{x}_k)=C$ for all $k$. The cyan and blue ellipses express the different values of $f(\mathbf{y}_k)$. The big red square $\mathbf{x}^*$ is contained in the left-inferior sub-rhombus, and the associated cyan point $\mathbf{y}_l$ in their center satisfy $f(\mathbf{y}_l) < f(\mathbf{y}_k)$ for all $k \neq l$.}
	\label{rhombus}
\end{figure}

In section III.A, we define the rhombus convergence property, Here, we provide the precise statement: Given a rhombus geometry defined by a non-singular matrix $\mathbf{A}$ and a vector $\mathbf{b}$, for a given rhombus with center $\mathbf{x}_0$ and  $2^N$ lattice vectors $\mathbf{y}_k$ (see Figure \ref{rhombus}), if the $\mathbf{x}^*$ point with $f(\mathbf{x}^*)=0$ is inside of the rhombus and between the points $\mathbf{y}_k$, the point $\mathbf{y}_1$ satisfy $f(\mathbf{y}_1)<f(\mathbf{y}_k)$ for all $k\neq 1$, then the point $\mathbf{x}^*$ is inside of the sub-rhombus associated to $\mathbf{y}_1$. Where, $f(\mathbf{x})=\Vert\mathbf{A}\cdot\mathbf{x}-\mathbf{b}\Vert$ and $\mathbf{A}\cdot\mathbf{x}^*=\mathbf{b}$. 

In order to understand this, we examine Figure \ref{rhombus}. Consider the optimal configuration $\mathbf{x}^*$ (big red square in the figure), an arbitrary ellipsoid of Figure \ref{FIG2}a centered in $\mathbf{x}^*$ and $2^N$ points $\mathbf{x}_k$'s defining the rhombus geometry (red ellipse and the four red squares in the figure). Consider another arbitrary point $\mathbf{x}_0$ (big blue point in the figure), which defines a new set of similar rhombus vectors $\mathbf{y}_k$'s (four cyan and blue points in the figure); for $N=2$, these rhombus vectors define the dashed rhombus, which is divided in $4$ sub-rhombus ($2^N$ in the arbitrary case). Suppose that $\mathbf{x}^*$ is inside of this rhombus and, therefore, also is inside of a particular sub-rhombus associated with the point $\mathbf{y}_l$ (cyan point in the left inferior corner of the figure), then, between the four $\mathbf{y}_k$'s the evaluated function $f(\mathbf{y}_l)$ reach the minimal value (in the figure, the cyan point in the left inferior corner is contained in the smaller cyan ellipse). We call this property the rhombus convergence, which is proved below. The property improves the convergence since the point associated with the QUBO solution $\mathbf{x}^*$ in each iteration is also contained in the next constructed rhombus.

Consider that the point $\mathbf{x}^*$ belongs to the sub-rhombus defined by $\mathbf{y}_1$. We prove that the point $\mathbf{y}_m$ that minimize the function $f(\mathbf{y}_k)$ restricted to the QUBO vectors $\mathbf{y}_k$`s satisfy $\mathbf{y}_k=\mathbf{y}_1$. As $\mathbf{x}^*$ belongs to the sub-rhombus defined by $\mathbf{y}_1$ we can write

\begin{equation}
	\mathbf{x}^*=\mathbf{y}_1 + \sum_{j=1}^{N}C_j\mathbf{v}_j, \,\,\,\, \mathrm{with}\,\, \vert C_j\vert \leq \frac{L}{4} \,\, \forall \,\, j,
\end{equation}
where $L$ is the side length of the principal rhombus shown in Figure \ref{rhombus} and $\mathbf{v}_j$ is the vector that defines the rhombus geometry. All points inside of the sub-rhombus associated with $\mathbf{y}_1$ satisfied $\vert C_j\vert \leq \frac{L}{4} \,\, \forall \,\, j,$ and any point outside of this sub-rhombus breaks the inequality. The point $\mathbf{x}^*$ also belongs to the principal rhombus, therefore
\begin{equation}\label{A1}
	\mathbf{x}^*=\mathbf{x}_0 + \sum_{j=1}^{N}D_j\mathbf{v}_j, \,\,\,\, \mathrm{with}\,\, \vert D_j\vert \leq \frac{L}{2} \,\, \forall \,\, j.
\end{equation}

It is clear that $\mathbf{x}^*-\mathbf{x}_0=\mathbf{x}_k-\mathbf{y}_k$, or
\begin{equation}\label{A2}
	\mathbf{y}_k=\mathbf{x}_k - \sum_{j=1}^{N}D_j\mathbf{v}_j.
\end{equation}

The function $f(\mathbf{y}_k)$ restricted to the points $\mathbf{y}_k$'s can be written as
\begin{equation}
	f(\mathbf{y}_k) = \Vert \mathbf{A}\cdot\left(\mathbf{x}^* - \mathbf{y}_k\right)\Vert^2,
\end{equation}
using eq.(\ref{A1}) and (\ref{A2}), we obtain
\begin{align}\label{A3}
f(\mathbf{y}_k)	=& \Vert\mathbf{A}\cdot(\sum_{j=1}^{N}D_j\mathbf{v}_j)\Vert^2 + \left\Vert\mathbf{A}\cdot(\mathbf{x}_k-\mathbf{x}^*)\right\Vert^2\nonumber \\& -2[\mathbf{A}\cdot(\sum_{j=1}^{N}D_j\mathbf{v}_j)]\cdot[\mathbf{A}\cdot(\mathbf{x}_k-\mathbf{x}^*)].
\end{align}
The two first terms of (\ref{A3}) are identical for all the possible $\mathbf{y}_k$ choices. Therefore, the minimal value of $f(\mathbf{y}_m)$ is reached by the $\mathbf{x}_m$ that maximize $$[\mathbf{A}\cdot(\sum_{j=1}^{N}D_j\mathbf{v}_j)]\cdot[\mathbf{A}\cdot(\mathbf{x}_m-\mathbf{x}^*)].$$

Consider a similar rhombus centered at $\mathbf{x}^*$ with associated rhombus points $\mathbf{x}_k$ (represented by red tiny squares in Figure \ref{rhombus}). The similar rhombus is not shown in the figure, but imagine it centered at $\mathbf{x}^*$. $\mathbf{x}_k$ belongs to this similar rhombus centered in $\mathbf{x}^*$ and is written as
\begin{equation}
	\mathbf{x}_k=\mathbf{x}^* + \sum_{j=1}^{N}s_j^{(k)}\frac{L}{4}\mathbf{v}_j,
\end{equation}
with $s_j^{(k)} \in \{-1,1\}$ for all $j$. Using the property $$(\mathbf{A}\cdot\mathbf{v}_i)\cdot(\mathbf{A}\cdot\mathbf{v}_j)=\mathbf{v}_i\cdot\left(\mathbf{A}^T\mathbf{A}\right)\cdot\mathbf{v}_j= h_i\delta_{ij},$$ we have
\begin{align}\label{A4}
	[\mathbf{A}\cdot(\sum_{j=1}^{N}D_j\mathbf{v}_j)]\cdot[\mathbf{A}\cdot(\mathbf{x}_k-\mathbf{x}^*)]=\sum_{j=1}^{N}s_j^{(k)}\frac{L}{4}h_jD_j.
\end{align}
To obtain the configuration that maximize (\ref{A4}) choose $s_j^{(m)}=\mathrm{Sign}(D_j)$ (the $h_i$ numbers are always positive). Using
\begin{equation}
	\Vert\mathbf{A}\cdot(\sum_{j=1}^{N}D_j\mathbf{v}_j)\Vert^2=\sum_{j=1}^{N}h_jD_j^2
\end{equation}
and
\begin{equation}
	\left\Vert\mathbf{A}\cdot(\mathbf{x}_m-\mathbf{x}^*)\right\Vert^2 = \sum_{j=1}^{N}h_j\frac{L^2}{16}
\end{equation}
we obtain
\begin{equation}\label{A5}
	f(\mathbf{y}_m)=\sum_{j=1}^{N}h_j\left(\vert D_j\vert - \frac{L}{4}\right)^2=\sum_{j=1}^{N}h_jE_j^2,
\end{equation}
or
\begin{equation}
	\Vert \mathbf{A}\cdot\left(\mathbf{x}^* - \mathbf{y}_m\right)\Vert^2 = \Vert\mathbf{A}\cdot(\sum_{j=1}^{N}E_j\mathbf{v}_j)\Vert^2
\end{equation}
implying in
\begin{equation}
	\mathbf{x}^*=\mathbf{y}_m + \sum_{j=1}^{N}E_j\mathbf{v}_j
\end{equation}
If we prove that $\vert E_j\vert \leq \frac{L}{4} \,\, \forall \,\, j$, then $E_j=C_j$ and $\mathbf{y}_m=\mathbf{y}_1$. From (\ref{A5}), we have
\begin{equation}
	E_j^2 = \left(\vert D_j\vert - \frac{L}{4}\right)^2
\end{equation}
or
\begin{equation}
	E_j^2 - \frac{L^2}{16}= \vert D_j\vert^2 - \frac{L}{2}\vert D_j\vert,
\end{equation}
but
\begin{equation}
	\vert D_j\vert \leq \frac{L}{2} \Rightarrow \vert D_j\vert^2 - \vert D_j\vert\frac{L}{2} \leq 0
\end{equation}
therefore
\begin{equation}
	E_j^2 - \frac{L^2}{16} \leq 0
\end{equation}
that is equivalent to $\vert E_j \vert \leq L/4$, hence $\mathbf{y}_m = \mathbf{y}_1$.


\section{Enhacement of algorithm 2 to construct the vectors $\mathbf{v}_k$}\label{RC2}

There is a result that improves the Algorithm 2 for calculating the vectors $\mathbf{v}_k$, with $k \in {1,\cdots,N}$. Suppose that we use the generalized Gram-Schmidt procedure $\mathbf{u}_k \to \mathbf{v}_k$, from $k=1$ until $k = m \leq N $. Therefore, the operator (in bra-ket notation):

\begin{equation}
	\mathbf{G}_m\cdot \mathbf{H}=\sum_{i=1}^{m}\left(\frac{1}{\langle \mathbf{v}_i \vert \mathbf{H} \vert \mathbf{v}_i \rangle}\vert \mathbf{v}_i \rangle \langle \mathbf{v}_i \vert\right)\cdot\mathbf{H},
\end{equation}
acts as the identity in the subspace generated by $\{\mathbf{u}_1,\cdots,\mathbf{u}_m\}$. In particular, when $m=N$, the operator $\mathbf{G}_N=\sum_{i=1}^{N}\left(\frac{1}{\langle \mathbf{v}_i \vert \mathbf{H} \vert \mathbf{v}_i \rangle}\vert \mathbf{v}_i \rangle \langle \mathbf{v}_i \vert\right)$ corresponds to the inverse of $\mathbf{H}$.

To prove the last assertion, let the operator $$\mathbf{C}_m=\mathbf{G}_m\cdot\mathbf{H}$$ be applied to all the vectors $\{\mathbf{u}_1,\cdots,\mathbf{u}_m\}$. Firstly

\begin{equation}
	\mathbf{C}_m\vert \mathbf{u}_1 \rangle = \sum_{i=1}^{m}\frac{1}{\langle \mathbf{v}_i \vert \mathbf{H} \vert \mathbf{v}_i \rangle}\vert \mathbf{v}_i \rangle \langle \mathbf{v}_i \vert\mathbf{H}\vert \mathbf{u}_1 \rangle
\end{equation}
but $\vert \mathbf{u}_1 \rangle =\vert \mathbf{v}_1 \rangle$, and $\langle \mathbf{v}_i \vert\mathbf{H}\vert \mathbf{u}_1 \rangle = 0$, if $i>1$. Therefore

\begin{equation}
	\mathbf{C}_m\vert \mathbf{u}_1 \rangle = \frac{1}{\langle \mathbf{v}_1 \vert \mathbf{H} \vert \mathbf{v}_1 \rangle}\vert \mathbf{v}_1 \rangle \langle \mathbf{v}_1 \vert\mathbf{H}\vert \mathbf{u}_1 \rangle=\vert \mathbf{u}_1 \rangle.
\end{equation}

The vectors $\vert \mathbf{v}_i \rangle$ are normalized in the algorithm, but it is clear that $\mathbf{G}_m$ does not depend on a particular normalization of $\vert \mathbf{v}_i \rangle$. Consider the unnormalized version of Eq. (\ref{GramVect}), in bra-ket notation:

\begin{equation}\label{B4}
	\vert \tilde{\mathbf{v}}_i \rangle =	\vert \mathbf{u}_i \rangle + \sum_{k=1}^{i-1}\beta_{ik}\vert \tilde{\mathbf{v}}_k \rangle
\end{equation}
with $\beta_{ik}=-\langle \tilde{\mathbf{v}}_k \vert\mathbf{H}\vert \mathbf{u}_i \rangle/\langle \tilde{\mathbf{v}}_k \vert\mathbf{H}\vert \tilde{\mathbf{v}}_k \rangle$. Suppose that $j\leq m$, we have

\begin{align}\label{B5}
	\mathbf{C}_m\vert \mathbf{u}_j \rangle &= \sum_{i=1}^{m}\frac{1}{\langle \tilde{\mathbf{v}}_i \vert \mathbf{H} \vert \tilde{\mathbf{v}}_i \rangle}\vert \tilde{\mathbf{v}}_i \rangle \langle \tilde{\mathbf{v}}_i \vert\mathbf{H}\vert \mathbf{u}_j \rangle \nonumber\\ &=\sum_{i=1}^{j-1}\frac{\langle \tilde{\mathbf{v}}_i \vert\mathbf{H}\vert \mathbf{u}_j \rangle}{\langle \tilde{\mathbf{v}}_i\vert \mathbf{H} \vert \tilde{\mathbf{v}}_i \rangle}\vert \tilde{\mathbf{v}}_i \rangle \nonumber \\ &\,\,\,\,\,\,\,\,\,+\frac{\langle \tilde{\mathbf{v}}_j \vert\mathbf{H}\vert \mathbf{u}_j \rangle}{\langle \tilde{\mathbf{v}}_j \vert \mathbf{H} \vert \tilde{\mathbf{v}}_j \rangle}\vert \tilde{\mathbf{v}}_j \rangle\nonumber \\ &\,\,\,\,\,\,\,\,\,+\sum_{i=j+1}^{m}\frac{\langle \tilde{\mathbf{v}}_i \vert\mathbf{H}\vert \mathbf{u}_j \rangle}{\langle \tilde{\mathbf{v}}_i \vert \mathbf{H} \vert \tilde{\mathbf{v}}_i \rangle}\vert \tilde{\mathbf{v}}_i \rangle.
\end{align}

From $\vert \tilde{\mathbf{v}}_j \rangle =	\vert \mathbf{u}_j \rangle + \sum_{k=1}^{j-1}\beta_{jk}\vert \tilde{\mathbf{v}}_k \rangle$, we have

\begin{equation}
	\langle \tilde{\mathbf{v}}_i \vert\mathbf{H}\vert \tilde{\mathbf{v}}_j \rangle =	\langle \tilde{\mathbf{v}}_i \vert\mathbf{H}\vert \mathbf{u}_j \rangle + \sum_{k=1}^{j-1}\beta_{jk}\langle \tilde{\mathbf{v}}_i \vert\mathbf{H}\vert \tilde{\mathbf{v}}_k \rangle
\end{equation}
if $i>j$, the previous equation imply in $\langle \tilde{\mathbf{v}}_i \vert\mathbf{H}\vert \mathbf{u}_j\rangle=0$. However, if $i=j$ we have $\langle \tilde{\mathbf{v}}_j \vert\mathbf{H}\vert \mathbf{u}_j\rangle=\langle \tilde{\mathbf{v}}_j \vert\mathbf{H}\vert \tilde{\mathbf{v}}_j\rangle$. Substituting in eq. (\ref{B5}):

\begin{align}
	\mathbf{C}_m\vert \mathbf{u}_j \rangle &=\sum_{i=1}^{j-1}\frac{\langle \tilde{\mathbf{v}}_i \vert\mathbf{H}\vert \mathbf{u}_j \rangle}{\langle \tilde{\mathbf{v}}_i\vert \mathbf{H} \vert \tilde{\mathbf{v}}_i \rangle}\vert \tilde{\mathbf{v}}_i \rangle \nonumber \\ &\,\,\,\,\,\,\,\,\,+\frac{\langle \tilde{\mathbf{v}}_j \vert\mathbf{H}\vert \tilde{\mathbf{v}}_j \rangle}{\langle \tilde{\mathbf{v}}_j \vert \mathbf{H} \vert \tilde{\mathbf{v}}_j \rangle}\vert \tilde{\mathbf{v}}_j \rangle,
\end{align}
but $\langle \tilde{\mathbf{v}}_i \vert\mathbf{H}\vert \mathbf{u}_j \rangle/\langle \tilde{\mathbf{v}}_i\vert \mathbf{H} \vert \tilde{\mathbf{v}}_i \rangle=-\beta_{ji}$ when $i<j$. Therefore

\begin{equation}
	\mathbf{C}_m\vert \mathbf{u}_j \rangle=-\sum_{i=1}^{j-1}\beta_{ji}\vert \tilde{\mathbf{v}}_i \rangle + \vert \tilde{\mathbf{v}}_j \rangle=\vert \mathbf{u}_j \rangle
\end{equation}
so, $\mathbf{C}_m\vert \mathbf{u}_j\rangle=\vert \mathbf{u}_j\rangle$ for all $j \leq m$, but $\{\mathbf{u}_1,\cdots,\mathbf{u}_m\}$ form a basis, therefore $\mathbf{C}_m$ acts as the identity in the subspace generated by the vectors $\{\mathbf{u}_1,\cdots,\mathbf{u}_m\}$, with $m \leq N$.

Equation (\ref{B4}) can be rewritten as:

\begin{equation}
	\vert \tilde{\mathbf{v}}_m \rangle =	\vert \mathbf{u}_m \rangle - \sum_{i=1}^{m-1}\left(\frac{1}{\langle \mathbf{v}_i \vert \mathbf{H} \vert \mathbf{v}_i \rangle}\vert \mathbf{v}_i \rangle \langle \mathbf{v}_i \vert\right)\cdot\mathbf{H}\vert \mathbf{u}_m \rangle
\end{equation}
or

\begin{equation}
	\vert \tilde{\mathbf{v}}_m \rangle = \left(\mathbb{I} - \sum_{i=1}^{m-1}\frac{1}{\langle \mathbf{v}_i \vert \mathbf{H} \vert \mathbf{v}_i \rangle}\vert \mathbf{v}_i \rangle \langle \mathbf{v}_i \vert \cdot\mathbf{H}\right) \vert\mathbf{u}_m \rangle,
\end{equation}
Due to our previous result, the action of the operation within the parentheses is straightforward. We define $\mathbf{h}_m=\mathbf{H} \cdot \vert \mathbf{u}_m \rangle$, which correspond to the $m$-th row of $\mathbf{H}$. In vector notation, the vector $\tilde{\mathbf{v}}_m$ is

\begin{equation}
	\tilde{\mathbf{v}}_m=\mathbf{u}_m - \mathbf{G}_{m-1}\cdot\mathbf{h}_m
\end{equation}
with $\mathbf{G}_{m-1}\cdot \mathbf{h}_m$ a vector with dimension equal to $m-1$.

\begin{figure}
	\centering
	\includegraphics[width=1\columnwidth]{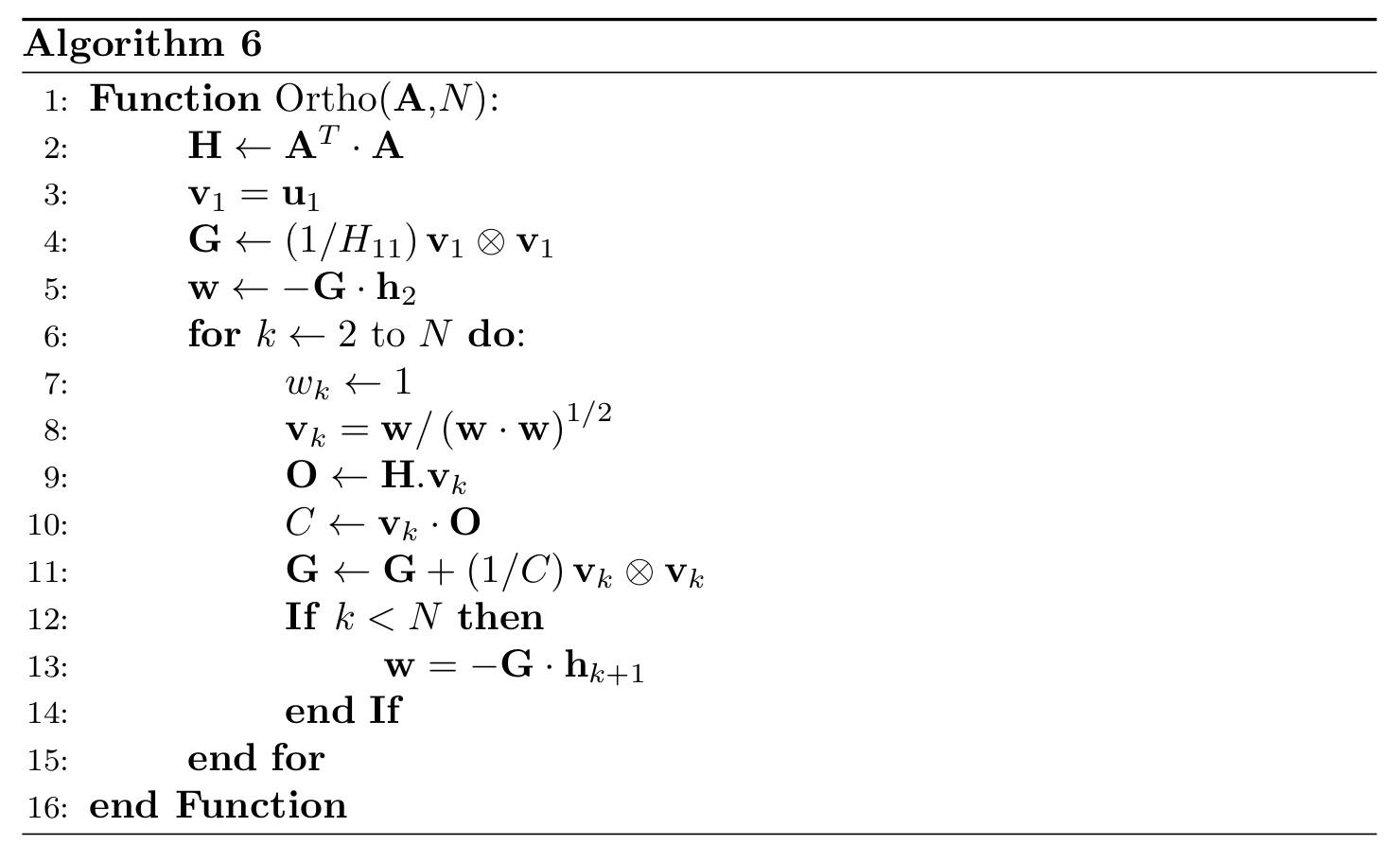}
	\caption{Gram-Schmidt procedure for the calculus of the $N$’s $\mathbf{H}$-orthogonal vectors $(\mathbf{v}_1,\cdots,\mathbf{v}_N)$. $w_k$ corresponds to the $k$-th vector component of $\mathbf{w}$ and $\mathbf{h}_k=\mathbf{H}\cdot \mathbf{u}_k$.}
	\label{algo6}
\end{figure}

The modification of algorithm 2 is shown in Figure \ref{algo6} 

To conclude, it is interesting to verify that for $m<N$, the operators

\begin{equation}
	\mathbf{E}_m=\sum_{i=1}^{m}\frac{1}{\langle \mathbf{v}_i \vert \mathbf{H} \vert \mathbf{v}_i \rangle}\vert \mathbf{v}_i \rangle \langle \mathbf{v}_i \vert \cdot\mathbf{H}
\end{equation}
and 
\begin{equation}
	\mathbf{E}_m^{c}=\mathbb{I} - \mathbf{E}_m
\end{equation}
are oblique non-Hermitian projectors. 

This is $$\mathbf{E}_m\cdot\mathbf{E}_m = \mathbf{E}_m,$$ 

$$\mathbf{E}_m^c\cdot\mathbf{E}_m^c = \mathbf{E}_m^c,$$

$$\mathbb{I}=\mathbf{E}_m+\mathbf{E}_m^c,$$
and
$$\mathbf{E}_m\cdot\mathbf{E}_m^c = \mathbf{E}_m^c\cdot\mathbf{E}_m=\mathbf{0}.$$
with $\mathbf{0}$ the zero matrix.

\section{Application of simple examples for algorithms 3 and 5}\label{app3}

In Section \ref{Sec1A}, we explicitly calculate the first iteration of Algorithm 1 shown in Fig. \ref{algo1} for a simple linear equation system with $N=2$. Here, we provide examples for two $4 \times 4$ ill-conditioned matrices applied to Algorithms 3 and 5 (see Fig. \ref{algo3} and Fig. \ref{algo5}), respectively. We use exact rational numerical expressions to demonstrate that the method works for arbitrary ill-conditioned matrices when the numerical error is controlled.

\subsection{Example for algorithm 3}

Consider the following ill-conditioned $4 \times 4$ matrix studied in Ref. \cite{Rump2009}.
\begin{widetext}
\begin{equation}
\mathbf{A} =	\left(\begin{matrix}
		-5046135670319638 && -3871391041510136 && -5206336348183639 && -6745986988231149\\
		-640032173419322 && 8694411469684959 && -564323984386760 && -2807912511823001\\
		-16935782447203334 && -18752427538303772 && -8188807358110413 && -14820968618548534\\
		-1069537498856711 && -14079150289610606 && 7074216604373039 && 7257960283978710
	\end{matrix}
	\right)
\end{equation}\label{Exact}
\end{widetext}

Next, we work with the exact rational expressions, but for simplicity and convenience, we express them in raw scientific notation between square brackets. This notation implies that we are working with involved exact rational expressions. For example, in this notation $\mathbf{A}$ is
\begin{equation}
			\left(\begin{matrix}
		-[5]\times 10^{15} && -[4]\times 10^{15} && -[5]\times 10^{15} && -[7]\times 10^{15}\\
		-[6]\times 10^{14} && [9]\times 10^{15} && -[6]\times 10^{14} && -[3]\times 10^{15}\\
		-[2]\times 10^{16} && -[2]\times 10^{16} && -[8]\times 10^{15} && -[1]\times 10^{16}\\
		-[1]\times 10^{15} && -[1]\times 10^{16} &&  [7]\times 10^{15} && [7]\times 10^{15}
	\end{matrix}
	\right) \nonumber
\end{equation}
means that we know the exact expression in (\ref{Exact}), but we do not show it explicitly. Define $\mathbf{H} = \mathbf{A}^T \cdot \mathbf{A}$, and apply Algorithm 2 or 6 to calculate the geometry vectors. It is convenient to exclude the step $ \mathbf{v}_k/(\mathbf{v}_k \cdot \mathbf{v}_k)^{1/2}\to \mathbf{v}_k$ to simplify the calculations. The geometry vectors are now not normalized, but the method still works. Defining $\mathbf{V} = \left(\mathbf{v}_1,\mathbf{v}_2,\mathbf{v}_3,\mathbf{v}_4\right)^T$ and $\mathbf{C}=\left(C_1,C_2,C_3,C_4\right)$ with $C_k=\mathbf{v}_k\cdot\left(\mathbf{H}\cdot\mathbf{v}_k\right)$, we obtain

\begin{equation}
	\mathbf{V}=\left(\begin{matrix}
		1 && 0 && 0 && 0\\
		-[1] && 1 && 0 && 0\\
		-[1] && [4]\times 10^{-1} && 1 && 	0\\
		-[8]\times 10^{-1} && [2]\times 10^{-1} &&  -[7]\times 10^{-1} && 1
	\end{matrix}
	\right), \nonumber
\end{equation}
and $\mathbf{C}=\left([3]\times 10^{32}, [3]\times 10^{32}, [2]\times 10^{31},[6]\times 10^{-97}\right).$

Next, we define $\mathbf{b}$ to apply Algorithm 3, choosing $$\mathbf{b} = \mathbf{A} \cdot \left(1, 20, 300, 4000\right)^T.$$

Therefore, at the final step of Algorithm 3, we must have $\mathbf{x}_0 \to \left(1, 20, 300, 4000\right)$. We explicitly show the results of the first iteration of the algorithm. Defining $\mathbf{I}=(1,1,1,1)$ and $\mathbf{A_q}=\mathbf{A}\cdot \mathbf{V}^T$ equal to

\begin{equation}
	\left(\begin{matrix}
		-[5]\times 10^{15} && [2]\times 10^{15} && -[2]\times 10^{15} && -[6]\times 10^{-49}\\
		-[6]\times 10^{14} && [9]\times 10^{15} && [4]\times 10^{15} && -[2]\times 10^{-49}\\
		-[2]\times 10^{16} && -[5]\times 10^{13} && [3]\times 10^{14} &&  [2]\times 10^{-49}\\
		-[1]\times 10^{15} && -[1]\times 10^{16} &&  [2]\times 10^{15} && -[2]\times 10^{-49}
	\end{matrix}
	\right) \nonumber
\end{equation}
Using the initial guess $\mathbf{x}_0 = (0, 0, 0, 0)$ and choosing $L = 10^5$, we calculate $$\mathbf{b_q} = \frac{1}{L}\left(\mathbf{b} + \frac{L}{2}\mathbf{A_q} \cdot \mathbf{I} - \mathbf{A} \cdot \mathbf{x}_0\right),$$
or $\mathbf{b_q}=\left(-[3]\times 10^{15}, [6]\times 10^{15}, -[9]\times 10^{15},-[5]\times 10^{15}\right).$

As we calculate the exact vector geometry of the problem, the QUBO matrix $\mathbf{Q}$ is diagonal. The diagonal is given by $$\mathbf{Q}=\mathbf{C}-2\mathbf{A_q}^T\cdot \mathbf{b_q}$$ or, $\mathbf{Q}=\left(-[2]\times 10^{31}, [1]\times 10^{31}, -[1]\times 10^{30},-[4]\times 10^{-98}\right).$
The configuration that minimizes the previous diagonal QUBO problem is $\mathbf{q} = (1, 0, 1, 1)$. Inserting this into $$\mathbf{x}_0 + L\mathbf{V}^T\cdot\left(\mathbf{q}-\frac{\mathbf{I}}{2}\right)$$
we obtain the new $$\mathbf{x}_0 \approx (18712.5, -18623.5, 14709.2, 50000),$$ we redefine $L/2 \to L$, and repeat the procedure as many times as necessary calculating news $\mathbf{b_q}$ and $\mathbf{Q}$. After 50 iterations, the error difference is $\vert \mathbf{x}^* - \mathbf{x}_0 \vert = 6.9\times 10^{-11}$, with $\mathbf{x}^* = (1,20,300,4000)$.

\subsection{Example for algorithm 5}

Now, consider the matrix $\mathbf{A}$
\\
\begin{widetext}
	\begin{equation}
		\mathbf{A} =	\left(\begin{matrix}
			-15000000000001 && 35000000000011 && -14999999999999 && 34999999999989\\
			35000000000011 && -15000000000001 && 34999999999989 && -14999999999999\\
			-14999999999999 && 34999999999989 && -15000000000001 && 35000000000011\\
			34999999999989 && -14999999999999 && 35000000000011 && -15000000000001
		\end{matrix}
		\right)
	\end{equation}\label{Exact2}
\end{widetext}  

Calculating $\mathbf{H} = \mathbf{A}^T\cdot \mathbf{A}$, call the first $2\times 2$ block diagonal block of $\mathbf{H}$ as $\mathbf{H}_1$.
$$\mathbf{H}_1= \left(\begin{matrix}
	[3]\times 10^{27} && -[2]\times 10^{27}\\
	-[2]\times 10^{27} && [3]\times 10^{27}
\end{matrix}\right)$$  
We can verify that the eigenvalues of $\mathbf{H}_1$ are $$E_{\mathbf{H}_1}=([5]\times 10^{27},[8]\times 10^{26}).$$ The square root of the quotient of these eigenvalues is 2.5, which is an excellent low condition number for a subproblem. Choose $(a_1, a_2) = (2, 2)$ and use Algorithm 4 to determine the vectors that decompose the original $4 \times 4$ problem into two $2 \times 2$ subproblems, in the notation of algorithm 4, $\mathbf{H}_I=\mathbf{H}_1^{-1}$. Also here, it is convenient to exclude the step $ \mathbf{v}_k/(\mathbf{v}_k \cdot \mathbf{v}_k)^{1/2}\to \mathbf{v}_k$ to simplify the calculations. In the end we obtain

\begin{equation}
	\mathbf{V} =\left(\begin{matrix}
		1 && 0 && 0 && 0\\
		0 && 1 && 0 && 0\\
		-[1] && [2]\times 10^{-25} && 1 && 	0\\
		[2]\times 10^{-25} && -[1] &&  0 && 1
	\end{matrix}
	\right), \nonumber
\end{equation}
and $$\mathbf{H_V}=\mathbf{V}\cdot\left(\mathbf{H}\cdot \mathbf{V}^T\right)=\left(\begin{matrix}
	\mathbf{H}_1 && \mathbf{0}_{2\times 2}\\
	\mathbf{0}_{2\times 2} && \mathbf{H}_2
\end{matrix}\right)$$
with $$\mathbf{H}_2= \left(\begin{matrix}
	[1]\times 10^{3} && -[2]\times 10^{2}\\
	-[2]\times 10^{2} && [1]\times 10^{3}
\end{matrix}\right).$$

As in the previous problem, define $$\mathbf{b} = \mathbf{A} \cdot \left(1, 20, 300, 4000\right)^T,$$ $L = 10^5$, and $\mathbf{x}_0 = (0, 0, 0, 0)$. Put $R=3$ and calculate $\mathbf{A_V}=\mathbf{A}\cdot\mathbf{V}^T$, $\mathbf{A_q} = \mathbf{A_V}\otimes \left(2^0,2^{-1},2^{-2}\right)$ and

$$\mathbf{I_q}^{(1)}=\mathbf{I_q}^{(2)} = \left(\begin{matrix}
	1 && 0\\
	0 && 1
\end{matrix}\right)\otimes \left(2^0,2^{-1},2^{-2}\right),$$
$$\mathbf{H_q}^{(j)} = \mathbf{H}_j\otimes \left(2^0,2^{-1},2^{-2}\right),$$
$$\mathbf{Q}_0^{(j)}=\left(\mathbf{I_q}^{(j)}\right)^T\cdot \mathbf{H_q}^{(j)},$$with $j \in \{1,2\}$.

With all these quantities, we can explicitly show the first iteration of Algorithm 5. First, we calculate $$\mathbf{b_q} = \frac{1}{L}\left(\mathbf{b}+L\frac{2^3-1}{2^3}\mathbf{A_V}\cdot \mathbf{I}- \mathbf{A}\cdot \mathbf{x}_0\right),$$ or $\mathbf{b_q} = ([2]\times 10^{13},[2]\times 10^{13},[2]\times 10^{13},[2]\times 10^{13})$. The dimension of $\mathbf{A_q}$ is $4 \times 12$, therefore $\mathbf{B} = \mathbf{A_q}^T \cdot \mathbf{b_q}$ has dimension $12$. The first $R * a_1 = 6$ components of $\mathbf{B}$ are associated with subproblem 1, and the remaining $6 = R * a_2$ components of $\mathbf{B}$ are associated with subproblem 2. Call such subvectors $\mathbf{B}_1$ and $\mathbf{B}_2$. The two QUBO subproblems to solve in the first iteration are $$\mathbf{Q}_j = \mathbf{Q}_0^{(j)} -2*\mathrm{Diag}^{(a_j)}\left(\mathbf{B}_j\right),$$ where $\mathrm{Diag}^{(a_j)}\left(\mathbf{B}_j\right)$ is a $6 \times 6 = R a_j \times R a_j$ diagonal matrix constructed with the vector $\mathbf{B}_j$ for $j \in {1, 2}$. Explicitly 

\begin{equation}
	\mathbf{Q}_1 =10^{26}\left(\begin{matrix}
[17] & [15] & [7.3] & -[21] & -[11] & -[5.3]  \\
[15] & [1] & [3.6] & -[11] & -[5.3] & -[2.6] \\
[7.3] & [3.6] & -[1.3]  &	-[5.3] & -[2.6]   & -[1.3]   \\
-[21]  & -[11] &  -[5.3]    &   [13]   & [15]  & [7.3]    \\ 
-[11] & -[5.3] & -[2.6] & [15] & -[0.85] & [3.6]\\ -[5.3] & -[2.6] & -[1.3] & [7.3] & [3.6] & -[2.2] 
	\end{matrix}\right), \nonumber
\end{equation}

\begin{equation}
	\mathbf{Q}_2 =10\left(\begin{matrix}
		-[42] & [49] & [24] & -[18] & -[8.8] & -[4.4]  \\
		[49] & -[45] & [12] & -[8.8] & -[4.4] & -[2.2] \\
		[24] & [12] & -[29]  &	-[4.4] & -[2.2]   & -[1.1]   \\
		-[18]  & -[8.8] &  -[4.4]    &   -[50]   & [49]  & [24]    \\ 
		-[8.8] & -[4.4] & -[2.2] & [49] & -[49] & [12]\\ -[4.4] & -[2.2] & -[1.1] & [24] & [12] & -[31] 
	\end{matrix}\right), \nonumber
\end{equation}
the best QUBO solutions are respectively $\mathbf{q}_1=\mathbf{q}_2=(1,0,0,1,0,0)$. We multiply each solution by the adequate factor as expressed in eq. \ref{NormVectors} and concatenate the two solutions in one. Explicitly
\begin{align}
\mathbf{q}_1 \to& (1\times 2^0,0\times 2^{-1},0\times 2^{-2},1\times 2^0,0\times 2^{-1},0\times 2^{-2}) \nonumber\\ \mathbf{q}_1^* = & (1,0,0,1,0,0) \nonumber \\
\mathbf{q}_2 \to& (1\times 2^0,0\times 2^{-1},0\times 2^{-2},1\times 2^0,0\times 2^{-1},0\times 2^{-2}) \nonumber\\ \mathbf{q}_2^* = & (1,0,0,1,0,0) \nonumber\\
\mathbf{q}= & (\mathbf{q}_1^*,\mathbf{q}_2^*)=(1,0,0,1,0,0,1,0,0,1,0,0)\nonumber
\end{align}
to obtain $\hat{\mathbf{x}}$, we add the four adjacent triplets ($R=3$ and $N=4$) of $\mathbf{q}$, explicitly
\begin{equation}
	\hat{\mathbf{x}} = (\underbrace{1,0,0}_{\mathrm{add}},\underbrace{1,0,0}_{\mathrm{add}},\underbrace{1,0,0}_{\mathrm{add}},\underbrace{1,0,0}_{\mathrm{add}})=(1,1,1,1).
\end{equation}
Inserting this into $$\mathbf{x}_0 + L\mathbf{V}^T\cdot\left(\hat{\mathbf{x}}-\frac{2^3-1}{2^3}\mathbf{I}\right)$$
we obtain the new $$\mathbf{x}_0 \approx (6.25\times10^{-21}, 6.25\times10^{-21}, 1250, 12500),$$
we redefine $L/2 \to L$, and repeat the procedure as many times as necessary calculating news $\mathbf{b_q}$, $\mathbf{Q}_1$ and $\mathbf{Q}_2$. After 120 iterations, the error difference is $\vert \mathbf{x}^* - \mathbf{x}_0 \vert = 2.0\times 10^{-32}$, with $\mathbf{x}^* = (1,20,300,4000)$.

\end{document}